\documentclass[aps,twocolumn,secnumarabic,groupedaddress,floatfix,amsfonts]{revtex4}
\usepackage[dvips]{graphicx,color}
\usepackage[dvips]{graphicx,color}
\begin{document}
\title{Parametrization dependence and $\mathbf{\Delta\chi^2}$ in parton 
distribution fitting}
\author{Jon Pumplin}
\affiliation{Michigan State University, East Lansing Michigan 48824, USA}
\date{\today}
\preprint{MSUHEP-090929}
\begin{abstract} 
Parton distribution functions, which describe probability densities 
of quarks and gluons in the proton, are essential for interpreting the data 
from high energy hadron colliders.  The parton distributions are measured by 
approximating 
them with functional forms that contain many adjustable parameters, which 
are determined by fitting a wide variety of experimental data. This paper 
examines the uncertainty that arises from choosing the form of parametrization, 
and shows how that uncertainty can be reduced using a technique based on 
Chebyshev polynomials.

\end{abstract} 
\maketitle
\section{Introduction
\label{sec:intro}}
Interpreting the data from high energy hadron colliders such as the Tevatron 
and LHC relies on parton distribution functions (PDFs), which describe the 
probability densities for quarks and gluons in the proton as a function of 
lightcone momentum fraction $x$ and QCD factorization scale $\mu$.  In current 
practice \cite{CT09,CT10,MSTW08,NNPDF}, the PDFs are measured by parametrizing 
them at a low scale $\mu_0$, using functional forms in $x$ that contain many
adjustable parameters.  The PDFs at higher $\mu$ are calculated 
using QCD renormalization group equations, and the best-fit parameter 
values are found through a ``global analysis,'' in which data from a 
variety of experiments are simultaneously fitted by minimizing a $\chi^2$ 
measure of fit quality.  

In the Hessian \cite{Hessian} and Lagrange multiplier \cite{Lagrange} methods, 
the uncertainty range of the PDFs is 
estimated by accepting all fits for which $\chi^2$ is not more than some fixed 
constant $\Delta\chi^2$ above the best-fit value.  Traditionally, 
$\Delta\chi^2 = 100$ \cite{cteq66} or $\Delta\chi^2 = 50$ \cite{MRST2004} have 
been used to estimate the 90\% confidence range.  
When these ``large'' $\Delta\chi^2$ values are used, weight factors 
or penalties may be included in the definition of the goodness-of-fit 
measure, to maintain an adequate fit to every individual data
set \cite{CT09}; or the uncertainty range along each eigenvector direction 
in the Hessian method can 
instead be estimated as the range where the fit to every experiment is 
acceptable \cite{MSTW08}.  
Large $\Delta\chi^2$ tolerance has long been a source of controversy.  Some 
groups \cite{HERA,Alekhin} reject it in 
favor of the $\Delta\chi^2=1$ for 68\% confidence ($\Delta\chi^2=2.7$ 
for 90\% confidence) that would be expected from Gaussian statistics;
however, results presented in this paper provide renewed evidence 
that substantially larger values of $\Delta\chi^2$ are necessary.

A potential motivation for large $\Delta\chi^2$ is based on conflicts
among the input data sets, which signal unknown systematic errors in the
experiments, or important theoretical errors introduced, \textit{e.g.}, by our 
reliance on leading-twist NLO perturbation theory, with a specific treatment
for heavy quarks.  It makes sense to scale up the experimental errors---which
is equivalent to raising $\Delta\chi^2$---to allow for such
conflicts \cite{PDG}.  However, the conflicts among experiments were recently
shown \cite{DSD,Consistency} to be fairly small:  the measured discrepancies
between each experiment and the collective implications from all of the
others suggest a minimum $\Delta\chi^2 \approx 10$ for 90\% confidence,
but supply no clear incentive for $\Delta\chi^2 \approx 100$.  
This is supported also by results from the NNPDF method \cite{NNPDF}, which 
finds conflicts among the data sets to be relatively small.  It is also 
supported by the distribution of $\chi^2$ per data point for the individual 
data sets \cite{CT10}; and by the observation that the average $\chi^2$ per 
data point in the global fit is close to 1, which suggests that the 
experimental errors are not drastically understated and that the theory 
treatment is adequate \cite{Lyons}.

Another source of uncertainty in PDF determination is the
\textit{parametrization dependence} error, which comes from representing 
the PDFs at $\mu_0$, which are unknown continuous functions, by expressions 
that are only adjustable through a finite number of free parameters.  
In traditional practice, flexibility is added to the parametrizing functions 
one parameter at a time, until the resulting minimum $\chi^2$ ceases to 
decrease ``significantly.'' However, at whatever point one chooses to stop 
adding fitting parameters, further small decreases in $\chi^2$ remain possible. 
This aspect of the PDF problem, namely that the number of fitting parameters 
is not uniquely defined, can spoil the normal rules, such as $\Delta\chi^2=1$ 
for 68\% confidence, which would otherwise follow from standard Gaussian 
statistics.  This point is illustrated in Sec.\ \ref{sec:example} by two 
hypothetical examples.  

A method that uses Chebyshev polynomials to dramatically increase the freedom 
of the parametrization, while maintaining an appropriate degree of smoothness 
in the resulting PDFs, is introduced in 
Sec.\ \ref{sec:chebyshev}.  
The Chebyshev method is applied to a typical PDF fit in 
Sec.\ \ref{sec:fits}.  
The method is further applied to the most recent CTEQ fit in 
Sec.\ \ref{sec:CT10}.  
Some aspects of the Chebyshev fit at large $x$ are discussed in 
Sec.\ \ref{sec:LargeX}.
Conclusions are presented in 
Sec.\ \ref{sec:conclusion}.

\section{Hypothetical examples
\label{sec:example}}
Let $z$ represent a displacement from the minimum point in $\chi^2$, along some 
specific direction in the space of fitting parameters.  It can be normalized 
such that 
\begin{equation}
\chi^2 \, = \, z^2 \, + \, C 
\label{eq:eq1}
\end{equation}
in the neighborhood of the minimum.
The parameter $z$ could be any one of the eigenvector coefficients $z_i$ that 
are discussed in \cite{DSD} or \cite{Consistency}.  Or by means of a suitable 
linear transformation $X = \alpha + \beta z$, $z$ could represent the prediction 
for some cross section $X$ that depends on the PDFs; or simply a PDF itself 
for some specific flavor, $x$, and $\mu$.  According to standard statistics, 
Eq.~(\ref{eq:eq1}) would imply $z = 0 \pm 1$ at 68\% confidence and $0 \pm 1.64$ 
at 90\% confidence.  If we assume instead---guided by \cite{Consistency}---that 
$\Delta\chi^2=10$ for 90\% confidence, we would expect $z = 0 \pm 3.16$ at that 
confidence.  However, the following argument shows that the uncertainty range 
may in principle be much broader than that. 

\begin{figure}[htb]
\vskip 20pt
\mbox{
 \resizebox*{0.22\textwidth}{!}{
\includegraphics[clip=true,scale=1.0]{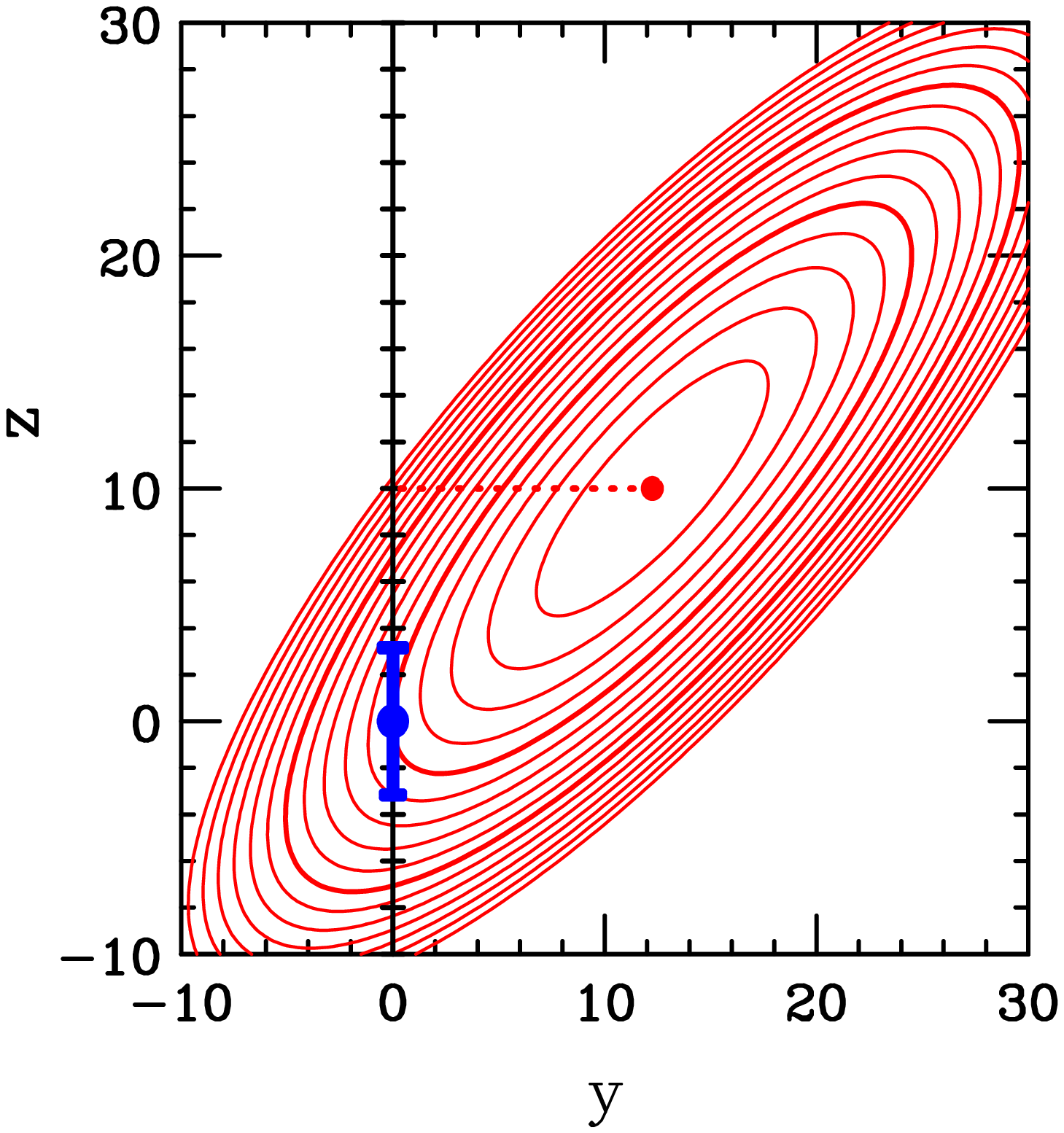}}
\hfill
\qquad
 \resizebox*{0.22\textwidth}{!}{
\includegraphics[clip=true,scale=1.0]{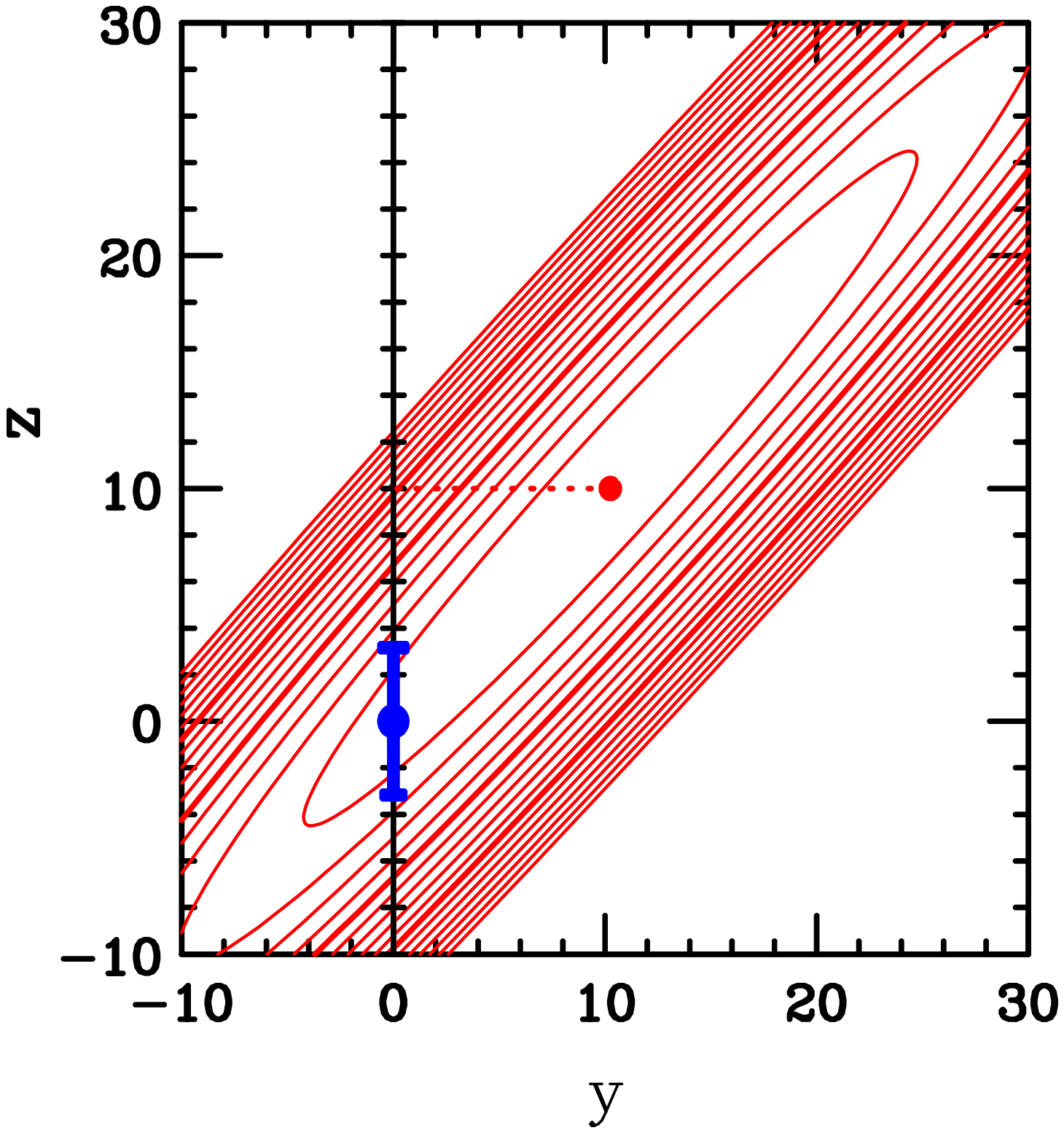}}
}
\vskip -10pt
 \caption{Contours of $\chi^2 = 3010$, $3020$, $3030,\dots$
in two hypothetical examples.  In each case, the best fit has 
$\chi^2=3000$ at $z = 10$. Meanwhile, 
if the fit is confined to $y=0$, the 
$\Delta\chi^2=10$ error limits appear to be $z = 0 \pm 3.16$, which 
is far too restrictive.
}
 \label{fig:figOne}
\vskip 10pt
\end{figure}

Suppose that, in order to reduce the dependence on the choice of 
parametrization, we introduce additional flexibility into the PDF model 
through a new parameter $y$, which is defined such that $\chi^2$ reduces 
to Eq.~(\ref{eq:eq1}) at $y=0$.  (To achieve a substantially improved 
fit, it will likely be necessary to increase the flexibility in more than 
one flavor, and therefore it will be necessary to introduce several new 
fitting parameters.  The parameter $y$ thus represents displacement in a 
direction defined by some particular \textit{linear combination} of 
several new and old parameters.)

Figure \ref{fig:figOne} shows two hypothetical contour plots for $\chi^2$ as a 
function of $y$ and $z$.  The contour interval is 10.  In each case, introducing 
the new parameter reveals that $z=10$ is a better estimate of the true value 
of $z$, so the prediction according to $y=0$, that $z = 0 \pm 3.16$ at 90\% 
confidence, is inaccurate.  In the scenario of the left panel, the additional 
freedom measured by $y$ has reduced the best-fit $\chi^2$ by $50$; while in 
the right panel, the reduction is only $5$---a change so small that 
one might easily have been content to mistakenly settle for $y=0$.  

In the hypothetical examples of Fig.\ \ref{fig:figOne}, $\Delta\chi^2 = 10$ 
yields an estimate of uncertainty for fits with $y=0$ that is far too 
narrow---even though in one case, the additional freedom only allows $\chi^2$ 
to be lowered by 5.  Appendix 1 shows that the qualitative form of the 
dependence of $\chi^2$ on $y$ and $z$ shown in Fig.\ \ref{fig:figOne} arises 
rather generally, whenever additional freedom is introduced into the 
parametrizations.  However, it remains to be seen whether such large 
quantitative changes actually arise in typical PDF fitting.  
A new parametrization method introduced in Sec.\ \ref{sec:chebyshev} will be 
used to answer that question in Secs.\ \ref{sec:fits} and \ref{sec:CT10}.

\section{{Chebyshev parametrizations}
\label{sec:chebyshev}}
In a recent typical PDF fit (CT09) \cite{CT09}, the gluon distribution 
was parametrized by
\begin{equation}
    x \, g(x,\mu_0) =  a_0 \, x^{a_1} \, (1-x)^{a_2} \, e^{p(x)} \label{eq:fofx}
\end{equation}
where
\begin{equation}
    p(x) = a_3 \sqrt{x} \, + \, a_4 x \, + \, a_5 x^2 \;. \label{eq:pexample}
\end{equation}
The same form was used---with different parameters of course---for the valence 
quark distributions $u_v = u - \bar{u}$ and $d_v = d - \bar{d}$, except that 
$a_3$ was set to $0$ in $d_v$, because that distribution is less 
constrained by data.

To provide greater flexibility in the parametrization, it would be natural 
to replace $p(x)$ by a general polynomial in
$\sqrt{x}\,$:
\begin{equation}
    p(x) = \sum_{j=1}^{n} b_j \, x^{j/2} \; . \label{eq:xseries}
\end{equation}
This form has several attractive features:
\begin{enumerate}
\item
The power-law dependence at $x \to 0$, with subleading terms suppressed by 
additional powers of approximately $x^{0.5}$, is expected from Regge theory. 
\item 
The power-law suppression in $(1-x)$ at $x \to 1$ is expected from 
spectator counting arguments.
\item
The exponential form $e^{p(x)}$ allows for the possibility of a large
ratio between the coefficients of the power-law behaviors at 
$x \to 0$ and $x \to 1$, without requiring large coefficients.  It also 
conveniently guarantees that $g(x)$ is positive definite---although that 
could in principle be an unnecessarily strong assumption, since the 
$\overline{\mbox{MS}}$ parton distributions are not directly observable, 
so it is only required that predictions for all possible \emph{cross sections} 
be positive.  
\item 
Restricting the order $n$ of the polynomial in Eq.~(\ref{eq:xseries}) can 
help to express the assumed smoothness of the parton distributions; although 
if $n$ is large, additional conditions must be imposed to prevent unacceptably 
rapid variations.
\end{enumerate}
The constraints on smoothness and limiting behavior at $x \to 0$ and $x \to 1$ 
are important. For without them, the momentum sum rule 
\begin{equation}
\sum_a \int_0^1 f_a(x,\mu) \, x \, dx \, = \, 1
\label{eq:momsum}
\end{equation} 
and the valence quark number sum rules 
\begin{equation}
\int_0^1 u_v(x,\mu) \, dx \, = \, 2 \;, \qquad 
\int_0^1 d_v(x,\mu) \, dx \, = \, 1 
\label{eq:numsum}
\end{equation} 
would have no power, because mildly singular contributions 
near $x=1$ in (\ref{eq:momsum}) or near $x=0$ in (\ref{eq:numsum}) 
could make arbitrary contributions to those integrals, without 
otherwise affecting any predictions. 

In past practice, only a small number of nonzero parameters $b_j$ have been 
retained in (\ref{eq:xseries}), 
as exemplified by the typical choice (\ref{eq:pexample}).  
The number of parameters can be increased to add flexibility, and thereby 
reduce the dependence on choice of parametrization.
However, that quickly runs into a technical difficulty: as more fitting 
parameters are included, the numerical procedure to find the minimum of 
$\chi^2$ becomes unstable, with large coefficients and 
strong cancellations arising in $p(x)$. The resulting best fits, 
if they can be found at all, contain implausibly rapid variations in 
the PDFs as a function of $x$.

This technical difficulty can be overcome by a method based on Chebyshev 
polynomials.  These polynomials have a long tradition in numerical analysis, 
although they have only recently begun to be applied to PDF studies \cite{Radescu}.  
The Chebyshev polynomials are defined---and conveniently calculated---by 
recursion:
\begin{eqnarray}
     T_0(y) &=& 1 \, , \quad 
     T_1(y) = y \nonumber \\
 T_{n+1}(y) &=& 2yT_n(y) - T_{n-1}(y) \; . 
\end{eqnarray}
Since $T_j(y)$ is a polynomial of order $j$ in $y$, the parametrization
(\ref{eq:xseries}) can be rewritten as
\begin{eqnarray}
    p(x) = \sum_{j=1}^{n} c_j \, T_j(y)  \label{eq:tseries} \; ,
\end{eqnarray}
where $y = 1-2\sqrt{x}$ conveniently maps the physical region $0 < x < 1$ to 
$\, -1 < y < 1\,$.

The parameters $c_1,\dots,c_n$ are formally equivalent to the parameters 
$b_1,\dots,b_n$; but
they are more convenient for fitting, because the requirement for smoothness 
in the input PDFs forces the $c_j$ parameters to 
be reasonably small at large order $j$.
This can be seen from the following property of the Chebyshev polynomials:
\begin{equation}
T_j(y) = \cos(j\theta) \quad \mbox{where} \quad y = \cos\theta \; .
\end{equation}
With the mapping $y = 1-2\sqrt{x}\,$,
$T_j(y)$ has extreme values of $\pm 1$ at the endpoints and at $j-1$ points 
in the interior of the physical region $0 < x < 1\,$. Chebyshev polynomials 
of increasingly large $j$\ thus model structure at an increasingly fine scale 
in $x$. 

Because the Chebyshev method provides so much flexibility in the parametrized 
input forms, there is a danger that it will produce fits with an 
unreasonable amount of fine structure in their $x$ distributions---potentially 
lowering $\chi^2$ in a misleading way by producing fits that match some of the 
statistical fluctuations in the data.  This difficulty can be 
overcome by defining an effective goodness-of-fit measure that is equal 
to the usual $\chi^2$ plus a penalty term that is based on a measure of the 
structure in the input distributions.  A particular way to include this 
``soft constraint'' is described in Appendix 2.  (The method is a major 
improvement over a method used to enforce smoothness in a preliminary 
version of this paper, which was based solely on the magnitudes of the 
coefficients $c_j$.)

With the Chebyshev method, it becomes possible to produce fits with three to 
four times as many free parameters than were tractable in previous PDF fitting.  
The method is applied in the next section to examine the parametrization error 
in a traditional fit.

\section{{Fits using the Chebyshev method}
\label{sec:fits}}
\begin{figure}[htb]
\vskip 20pt
\mbox{
 \resizebox*{0.23\textwidth}{!}{
\includegraphics[clip=true,scale=1.0]{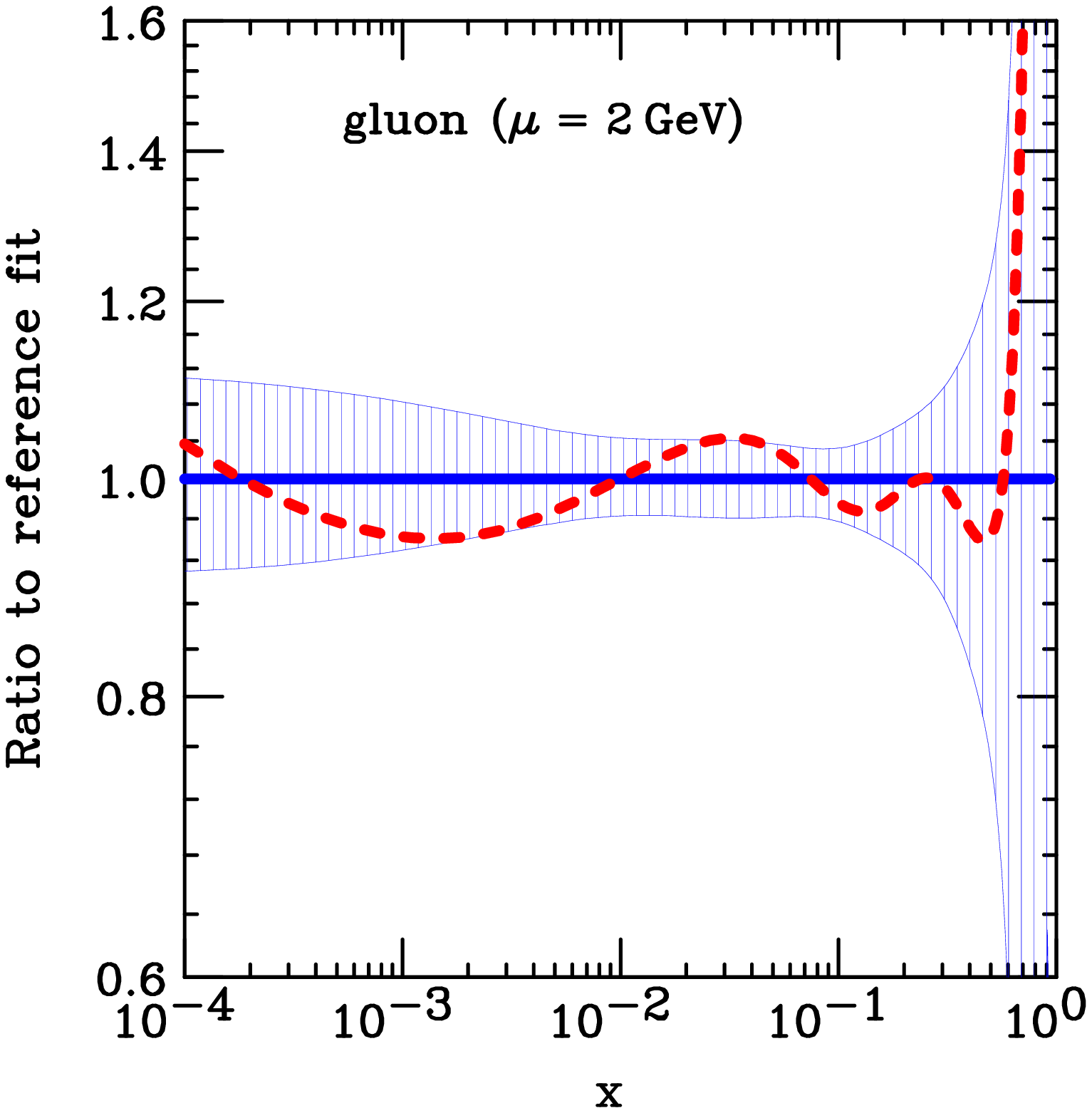}}
\hfill
 \resizebox*{0.23\textwidth}{!}{
\includegraphics[clip=true,scale=1.0]{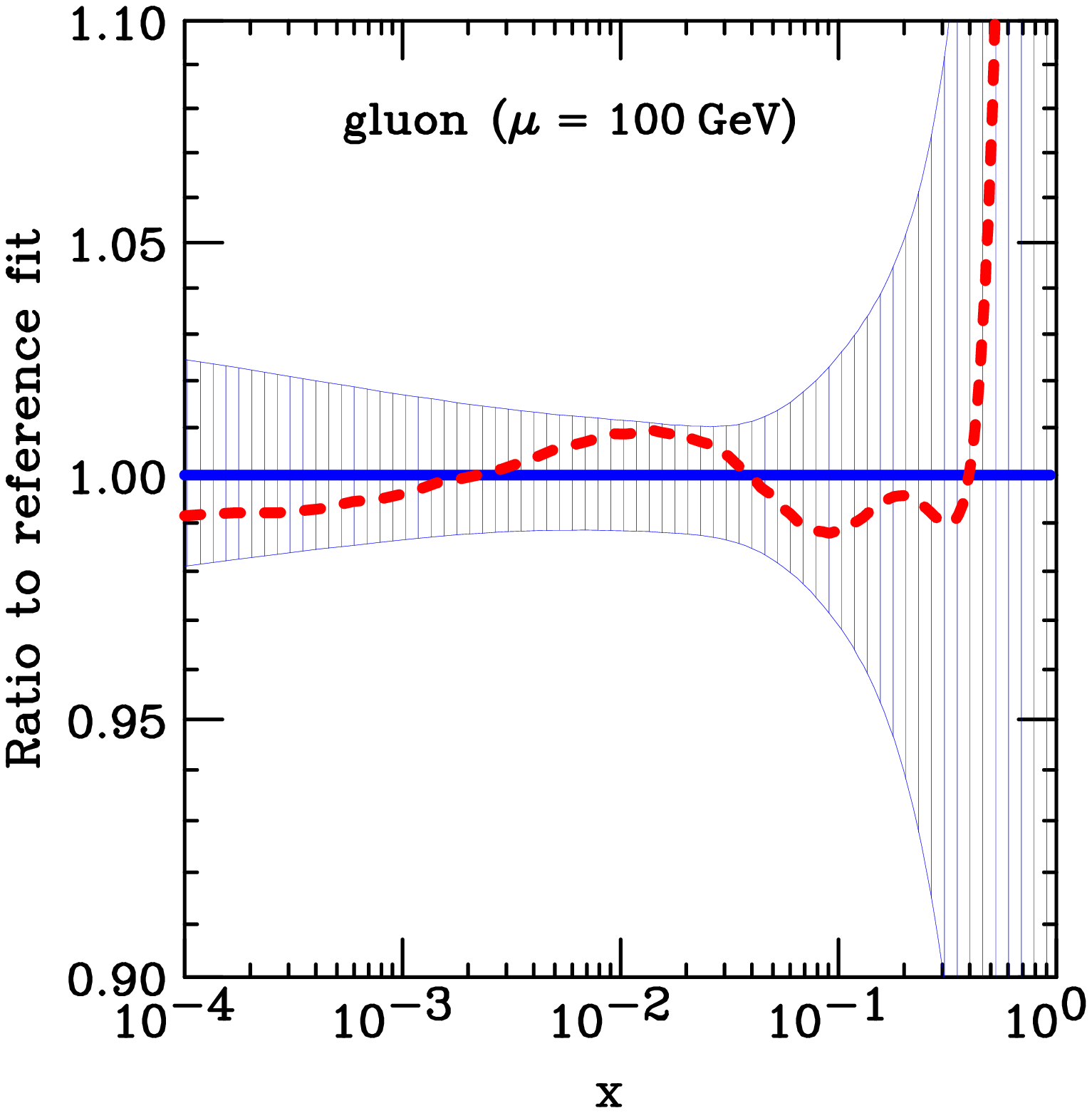}}
}
\mbox{
 \resizebox*{0.23\textwidth}{!}{
\includegraphics[clip=true,scale=1.0]{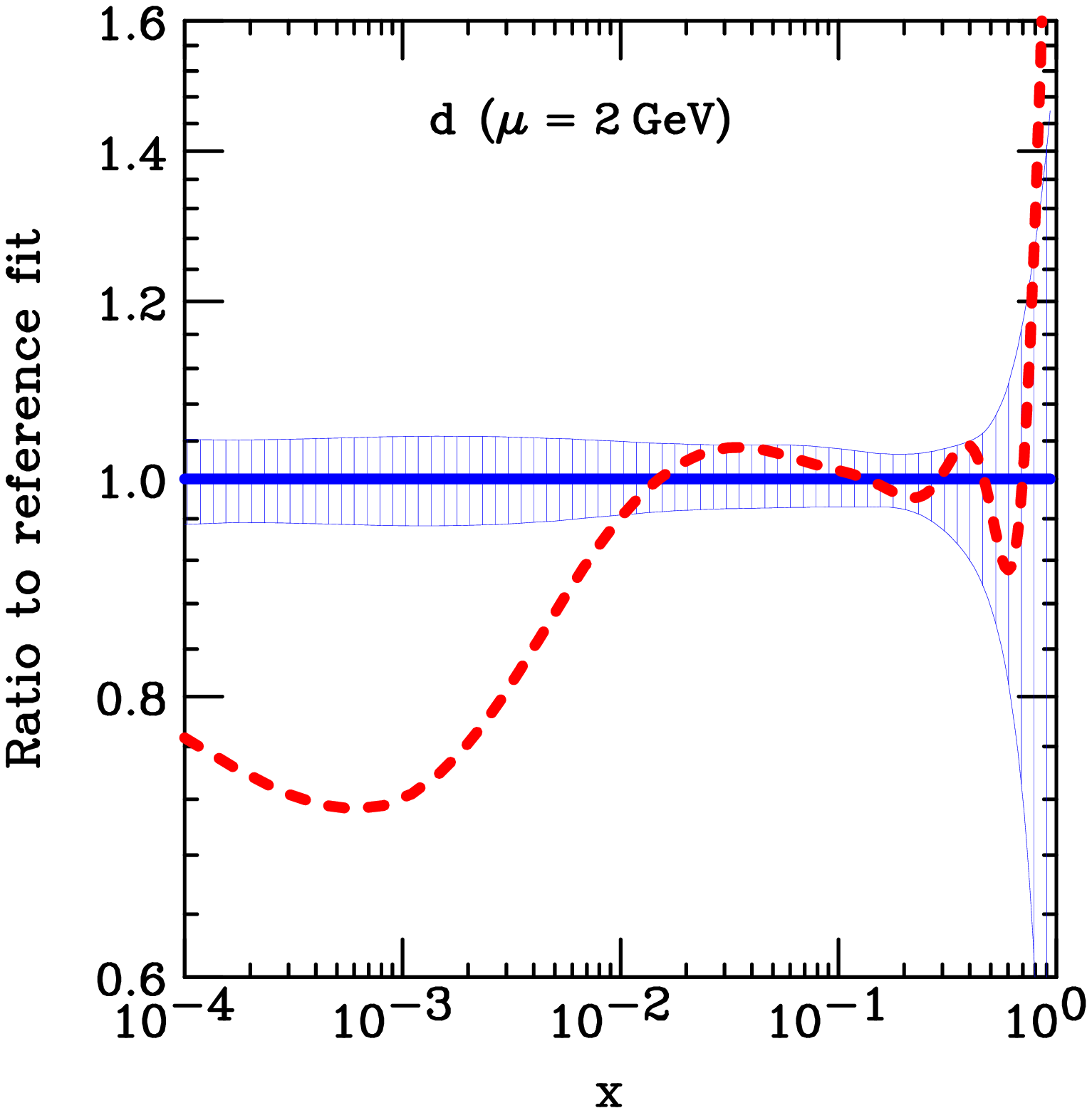}}
\hfill
 \resizebox*{0.23\textwidth}{!}{
\includegraphics[clip=true,scale=1.0]{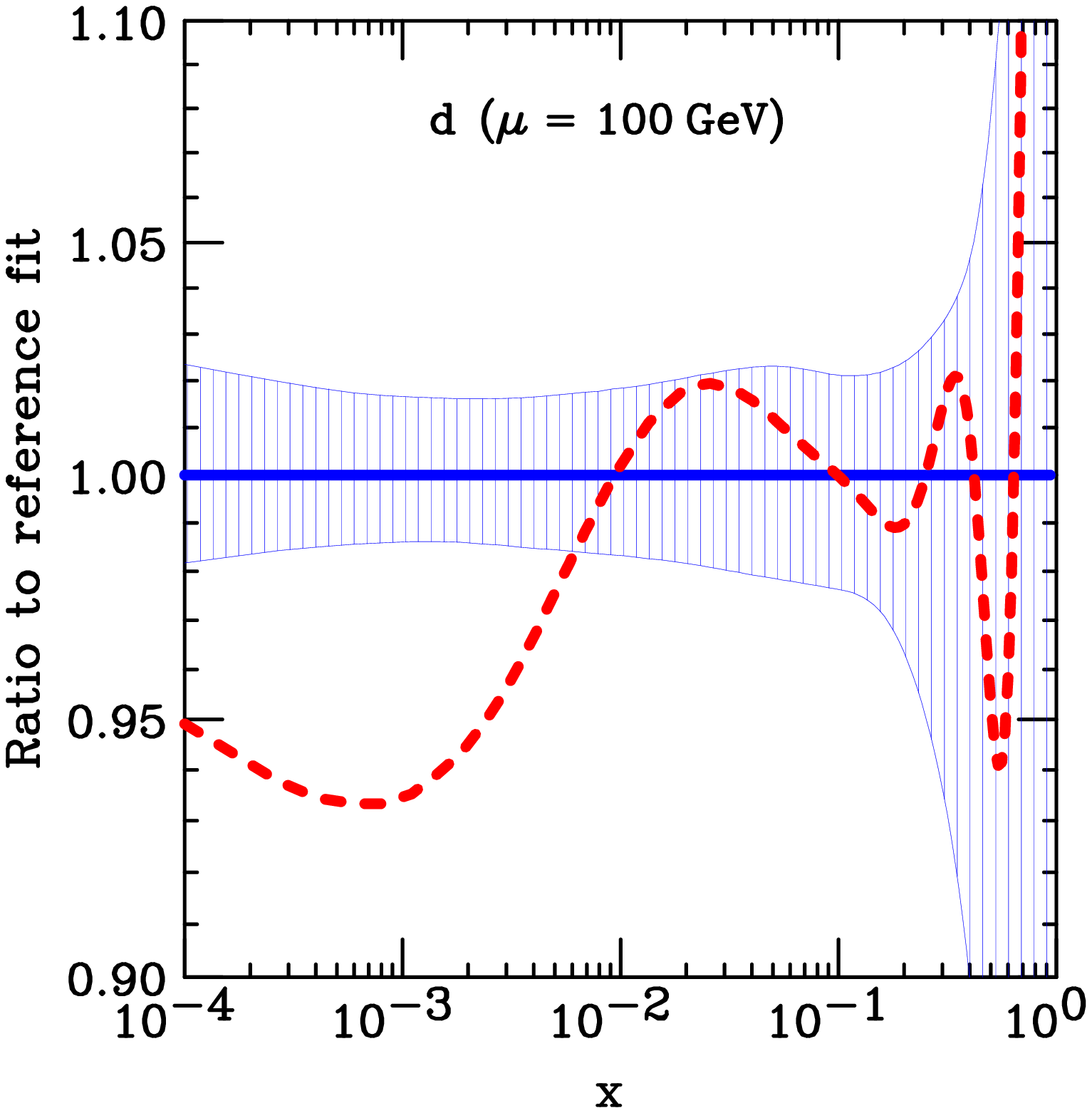}}
}
\mbox{
 \resizebox*{0.23\textwidth}{!}{
\includegraphics[clip=true,scale=1.0]{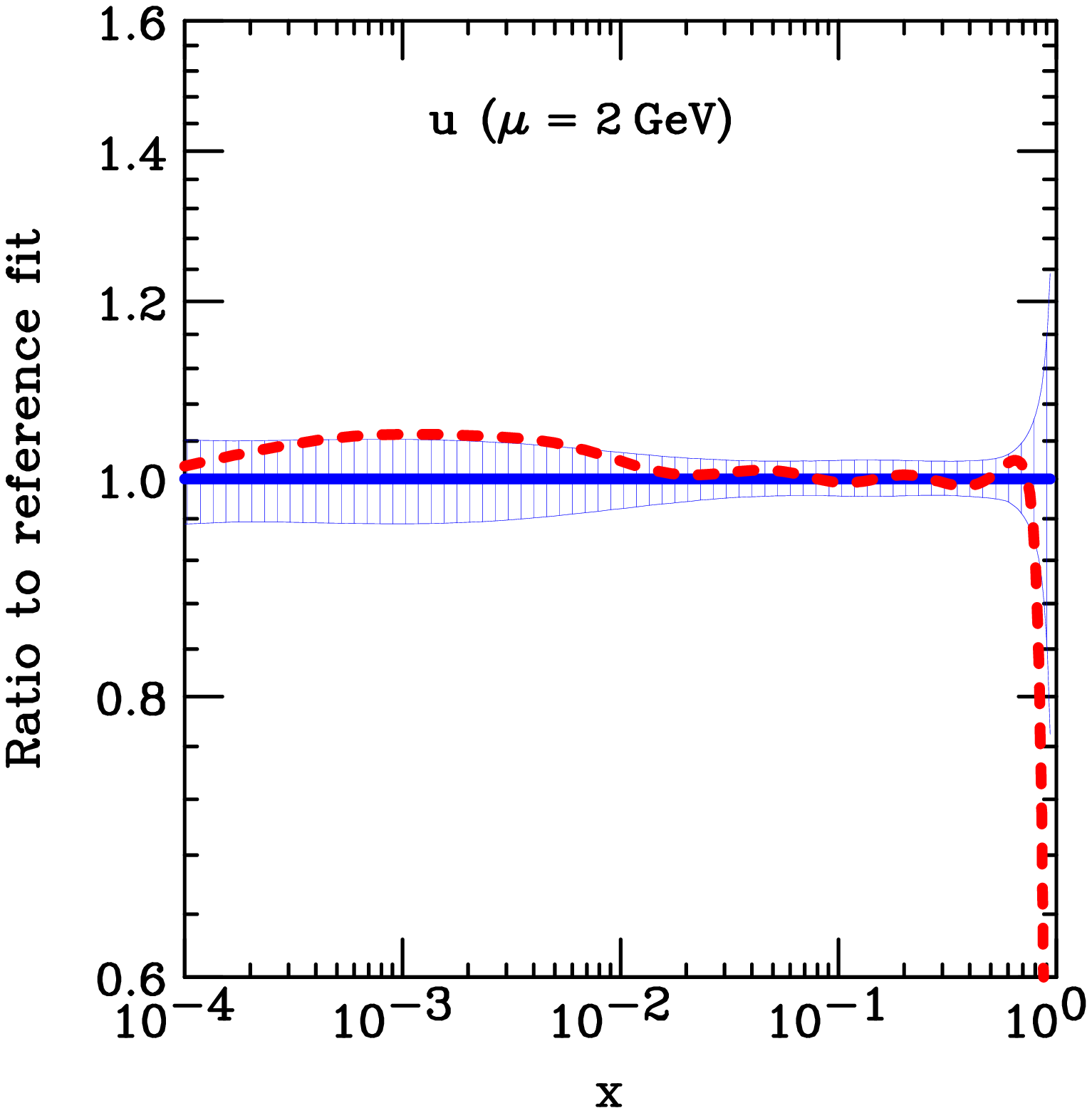}}
\hfill
 \resizebox*{0.23\textwidth}{!}{
\includegraphics[clip=true,scale=1.0]{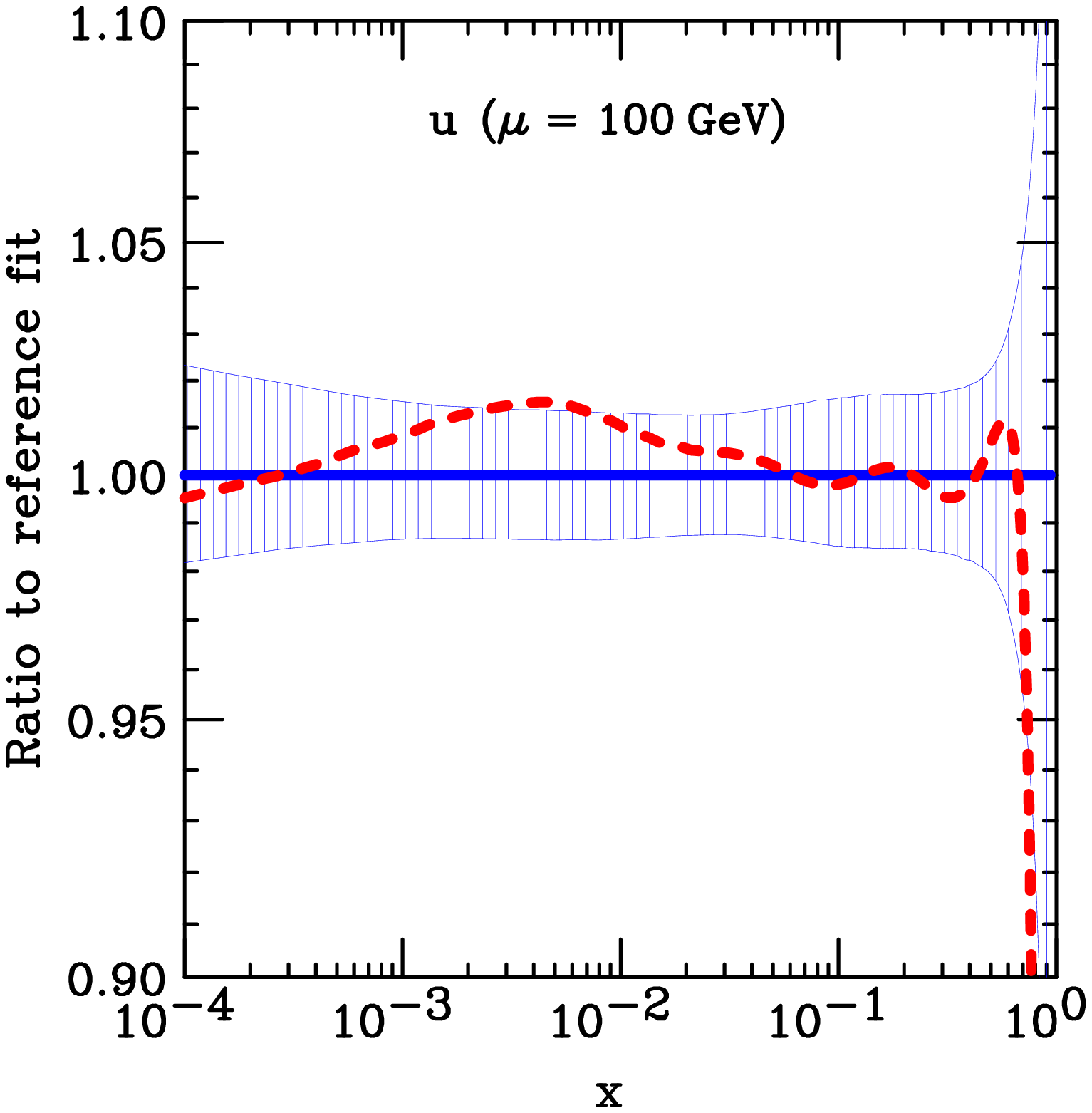}}
}
\vskip -10pt
 \caption{Fractional uncertainty of gluon distribution, and $d$ and $u$ 
quark distributions, calculated at $\Delta\chi^2 = 10$ using the CTEQ6.6 
form of parametrization with 22 parameters. Dashed curves are from a 
fit using the Chebyshev polynomial method with 71 parameters.  
}
 \label{fig:figTwo}
\vskip 10pt
\end{figure}
Figure \ref{fig:figTwo} shows the fractional uncertainty obtained using the 
parametrization method of CTEQ6.6 \cite{cteq66}, which has 22 free parameters.  
The uncertainty limit is defined here by $\Delta \chi^2 = 10$, which is the
range suggested by observed conflicts among the input data 
sets \cite{Consistency}.  The dashed curve shows
the result of a fit using the Chebyshev method described in Appendix 2. 
This fit has 71 free parameters, and achieves a $\chi^2$ that is lower by $72\,$.

We see that introducing the more flexible parametrization has shifted the 
best-fit estimate of the PDFs by an amount that is in a number of places 
comparable to the previously estimated $\Delta\chi^2=10$ uncertainty.
In some regions, namely at very large or very small $x$, the shift produced
by the change in parametrization is much larger than the previous 
uncertainty estimate.  
These are regions where the available data 
provide little constraint on the PDFs, so their estimated uncertainty 
in the CTEQ6.6-style fit was artificially small due to the lack of 
flexibility in that parametrization.  Other contemporary PDF fits 
use still less flexible parametrizations, which must underestimate 
the true uncertainty in those regions even more.

The fits shown in this section were made using a relatively crude method for 
imposing smoothness on the Chebyshev polynomial fits (based on limiting the 
magnitudes of the coefficients of those polynomials).  As a consequence, 
substantial deviations from $u(x) \approx d(x)$ appear at $x$ as small 
as $10^{-4}$, in spite of the limiting condition $u(x)/d(x) \to 1$ at 
$x \to 0$ that is assumed in all of these fits.  This can be seen in 
Fig.\ \ref{fig:figTwo}, where $u(x)$ and $d(x)$ are shifted quite 
differently from the CT10 reference fit, for which $u(x) \approx d(x)$ 
is a good approximation at small $x$.  An improved method for imposing 
smoothness is used in the fits shown in the next section, and that restores 
the $u(x) \approx d(x)$ behavior at small $x$. 

The second set of graphs in Fig.\ \ref{fig:figTwo} shows that parametrization 
effects are still important at the relatively large scale of 
$\mu = 100 \, \mathrm{GeV}$.  Hence they are an important source of uncertainty 
for many processes of interest at the Tevatron and LHC.

\section{{Application to CT10}
\label{sec:CT10}}
While this paper was being revised, an updated version of the CTEQ/TEA parton 
distribution analysis was completed.  This CT10 \cite{CT10} analysis includes 
a number of improvements to the previous CTEQ6.6 \cite{cteq66} and
CT09 \cite{CT09} analyses, and is now the most up-to-date of the CTEQ
PDF fits. It is therefore interesting to examine the uncertainty caused by 
parametrization dependence in this new fit, using the Chebyshev technique.

\begin{figure}[htb]
\vskip 20pt
\mbox{
 \resizebox*{0.23\textwidth}{!}{
\includegraphics[clip=true,scale=1.0]{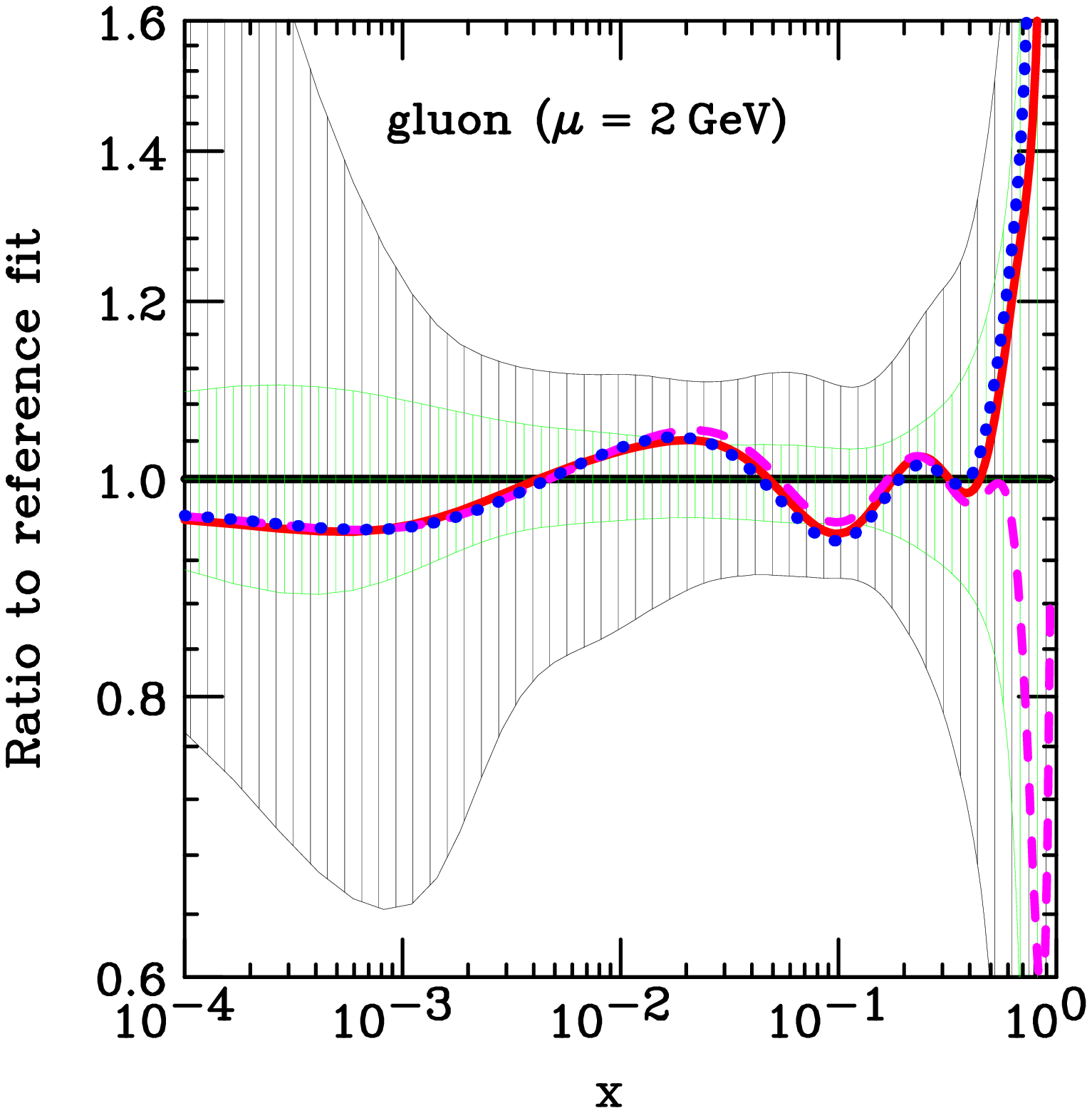}}
\hfill
 \resizebox*{0.23\textwidth}{!}{
\includegraphics[clip=true,scale=1.0]{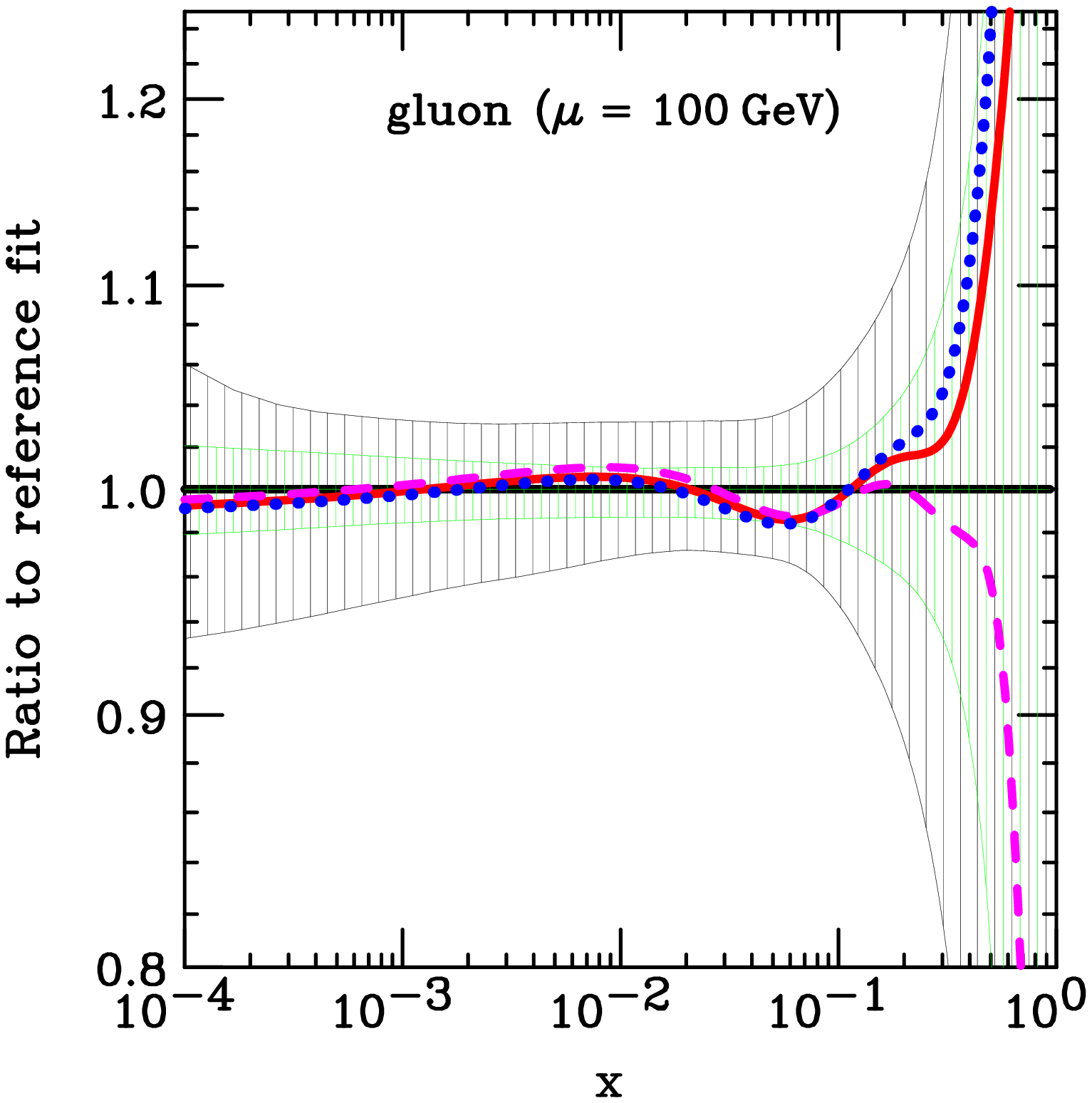}}
}
\mbox{
 \resizebox*{0.23\textwidth}{!}{
\includegraphics[clip=true,scale=1.0]{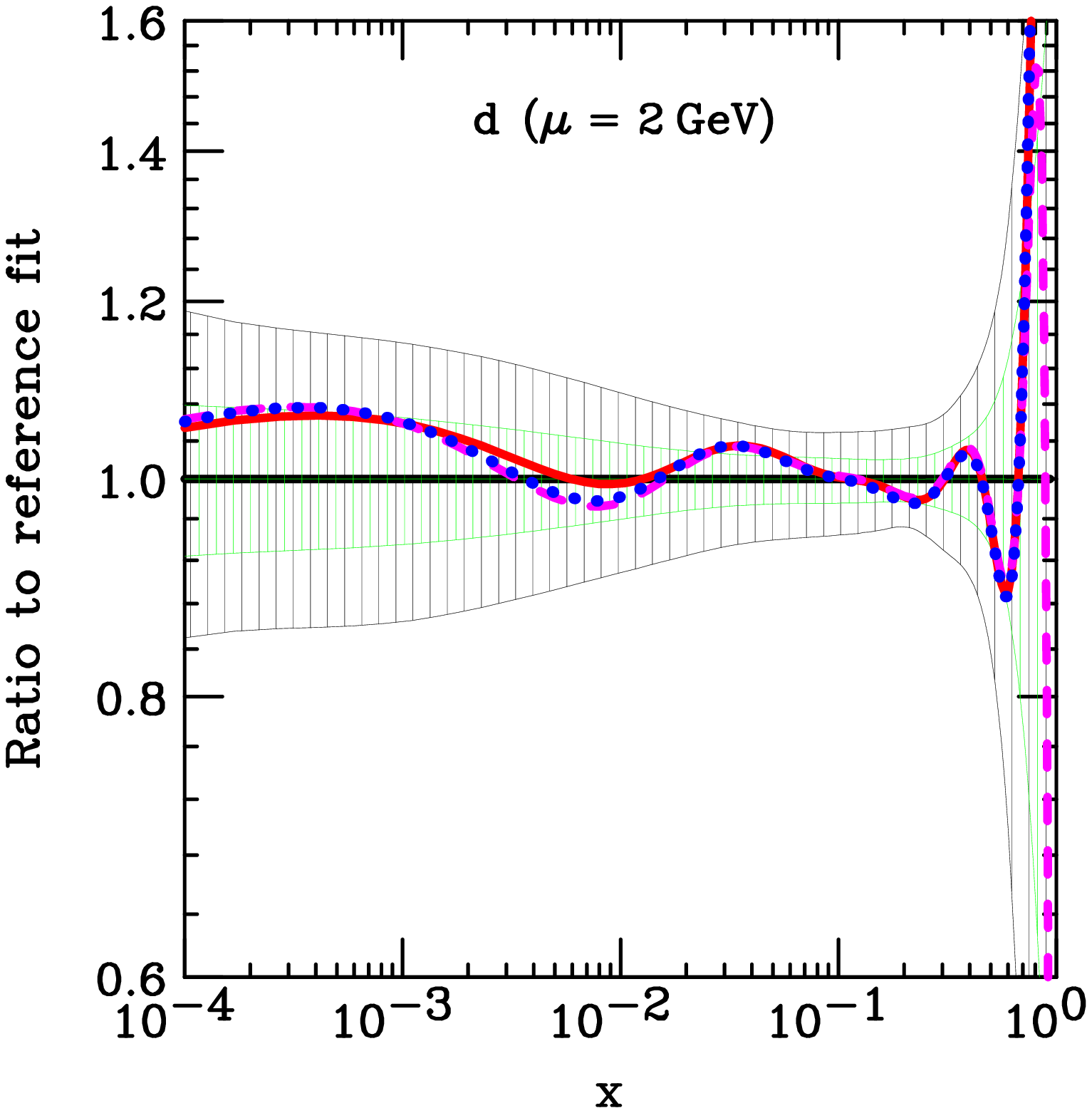}}
\hfill
 \resizebox*{0.23\textwidth}{!}{
\includegraphics[clip=true,scale=1.0]{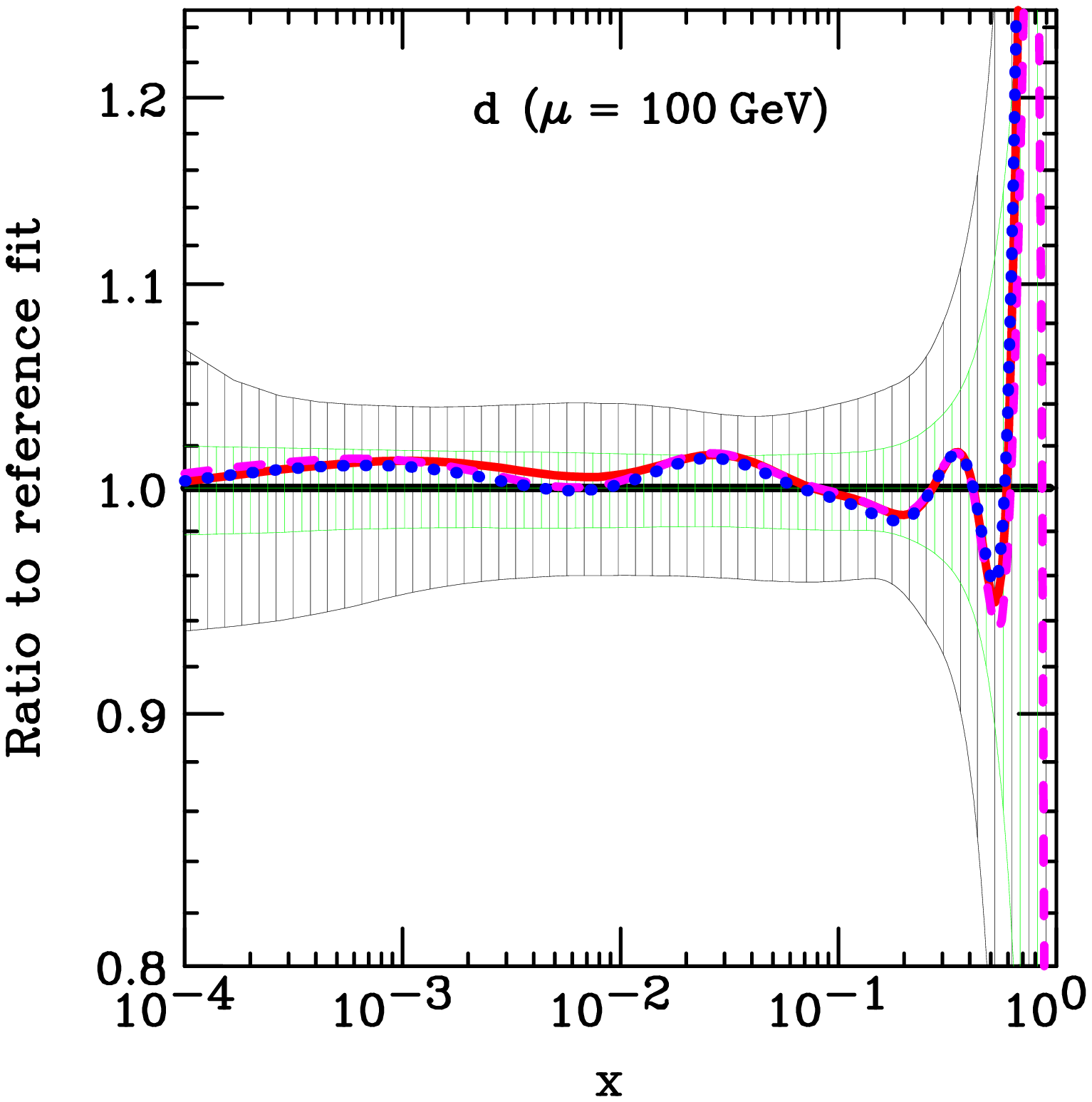}}
}
\mbox{
\hfill
 \resizebox*{0.23\textwidth}{!}{
\includegraphics[clip=true,scale=1.0]{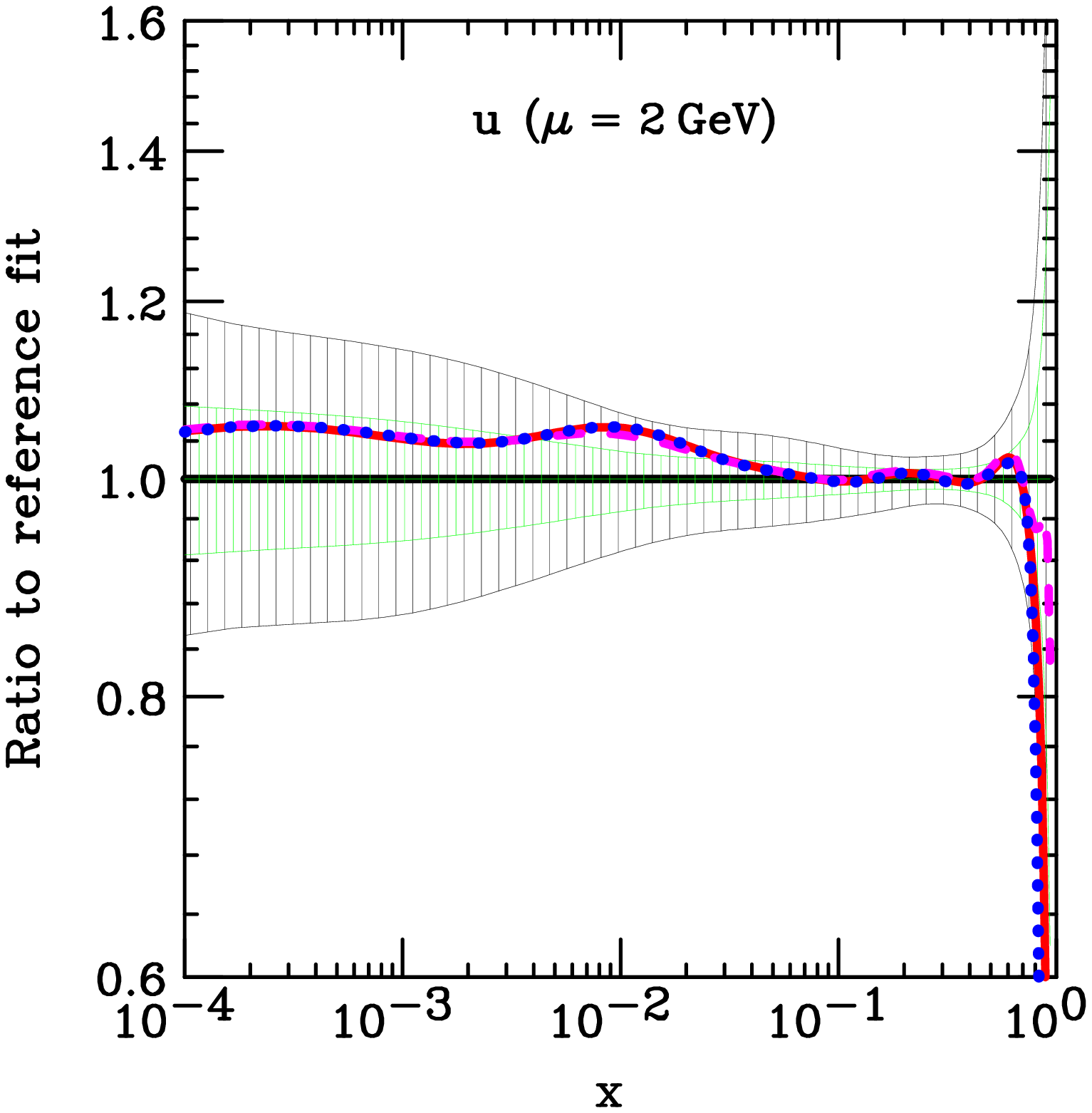}}
\hfill
 \resizebox*{0.23\textwidth}{!}{
\includegraphics[clip=true,scale=1.0]{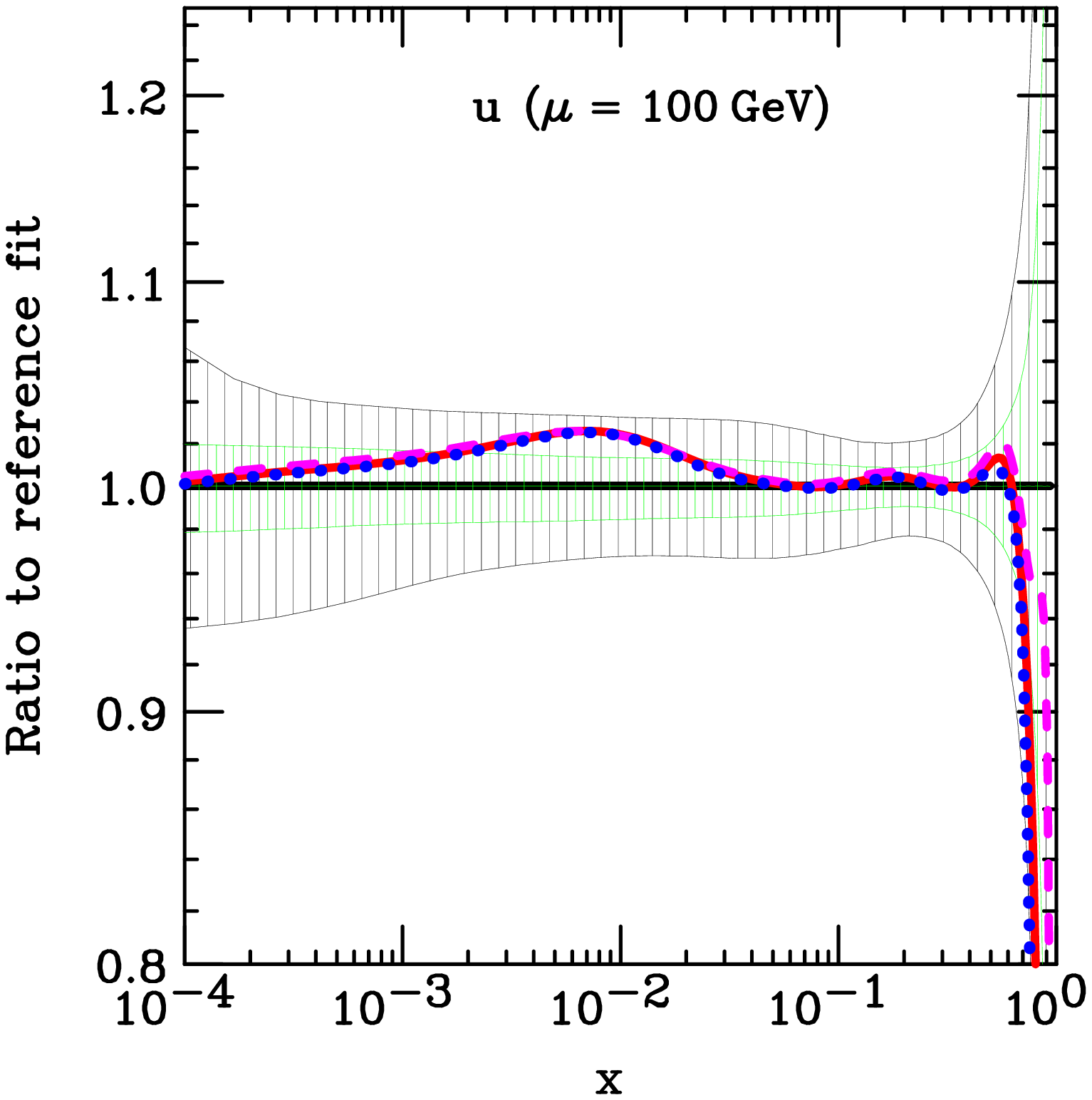}}
}
\vskip -10pt
 \caption{Wide shaded regions show the fractional uncertainty of gluon, 
$d$-quark, and $u$-quark distributions 
at scales $\mu = 2 \, \mathrm{GeV}$ and $\mu = 100 \, \mathrm{GeV}$ 
according to CT10 (26 fitting parameters). 
Narrow shaded regions show the corresponding uncertainty 
defined by a simple $\Delta\chi^2 = 10$ criterion.
The solid curves are from a fit using the Chebyshev polynomial method: 
this fit has 84 free parameters and its $\chi^2$ is lower than CT10 by 
$105$.  The dashed and dotted curves show similar Chebyshev fits with 
different behaviors at large $x$.
}
 \label{fig:figThree}
\vskip 10pt
\end{figure}
\begin{figure}[htb]
\vskip 20pt
\mbox{
 \resizebox*{0.21\textwidth}{!}{
\includegraphics[clip=true,scale=1.0]{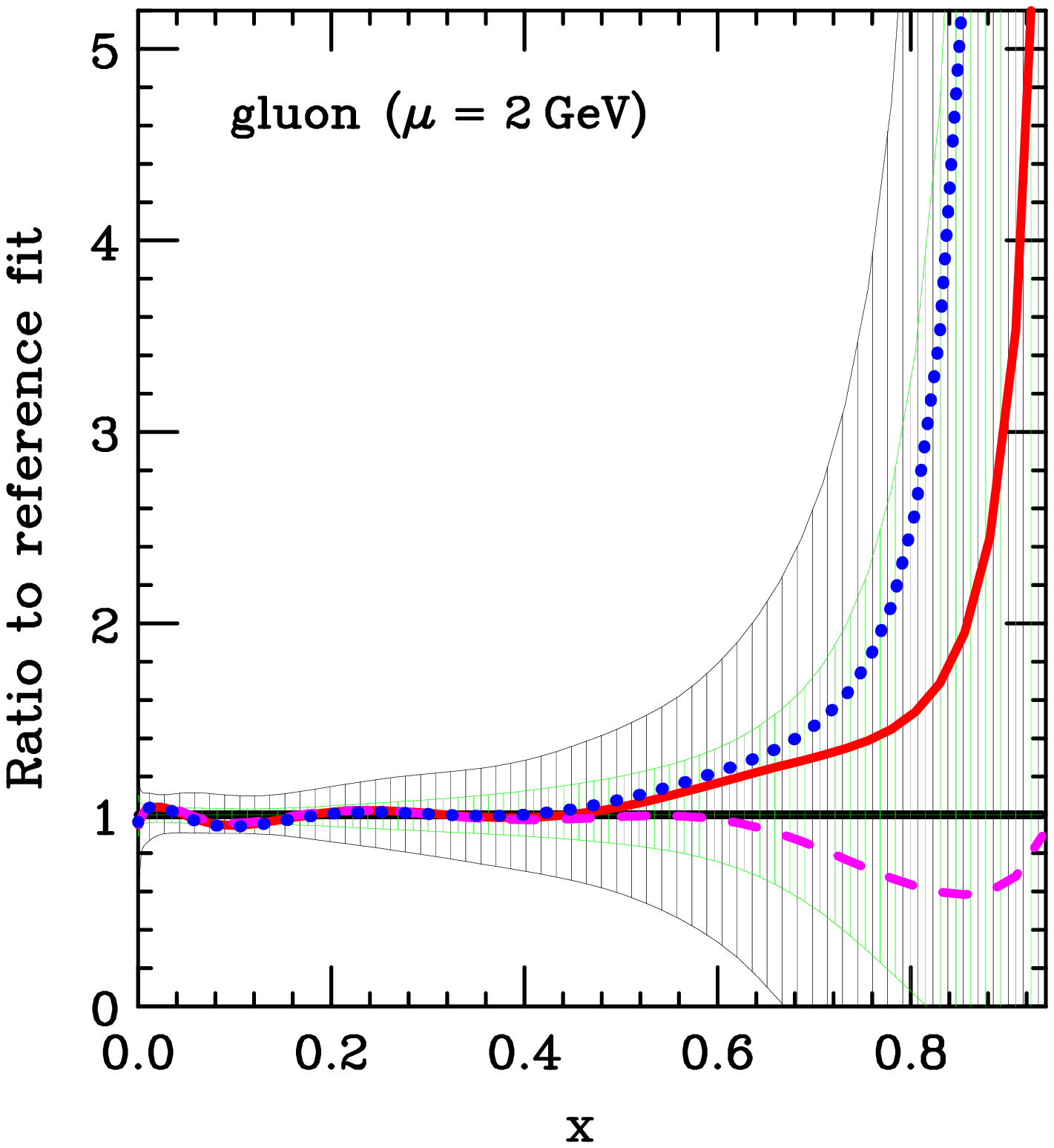}}
\hfill
 \resizebox*{0.21\textwidth}{!}{
\includegraphics[clip=true,scale=1.0]{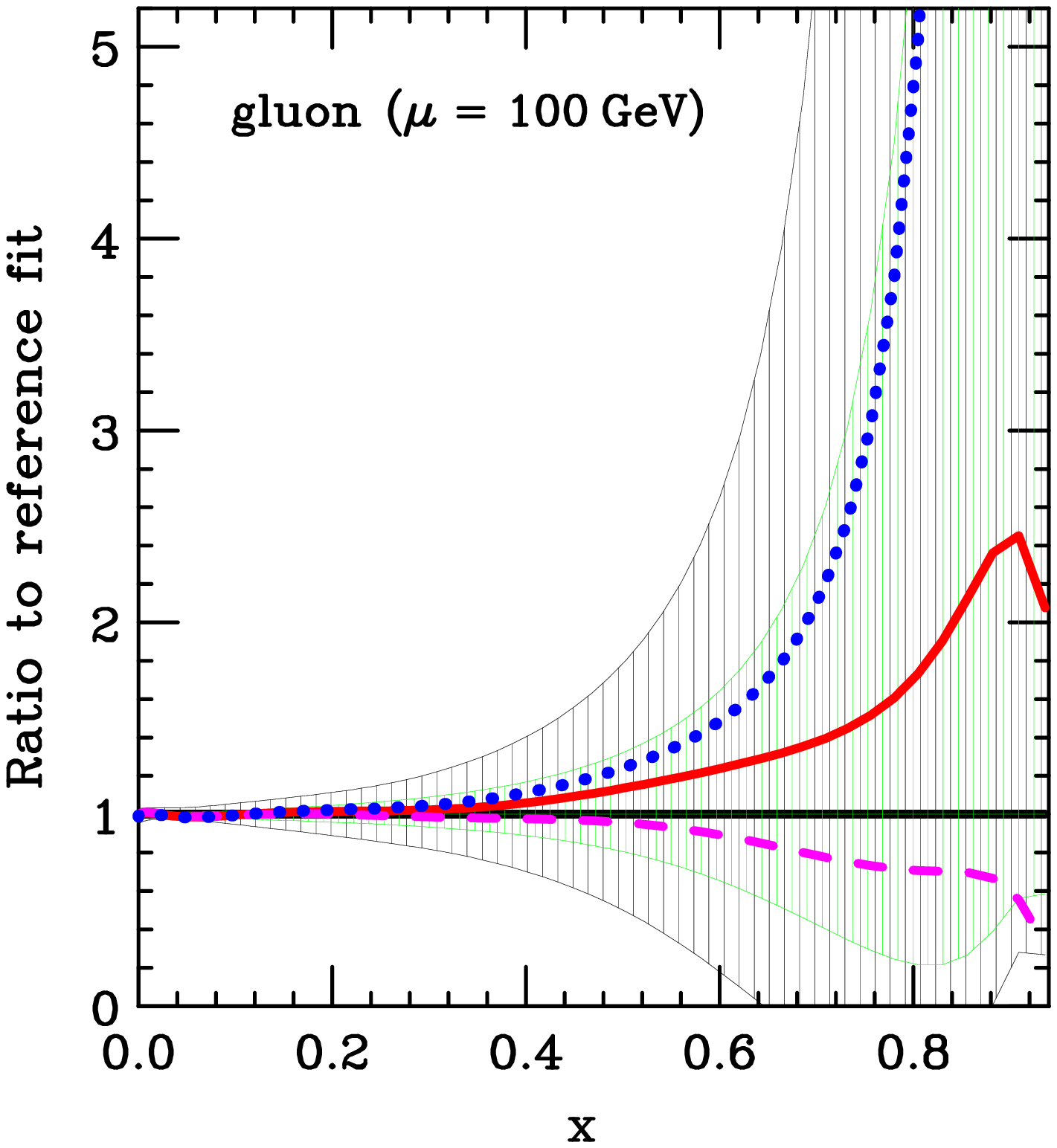}}
}
\mbox{
 \resizebox*{0.21\textwidth}{!}{
\includegraphics[clip=true,scale=1.0]{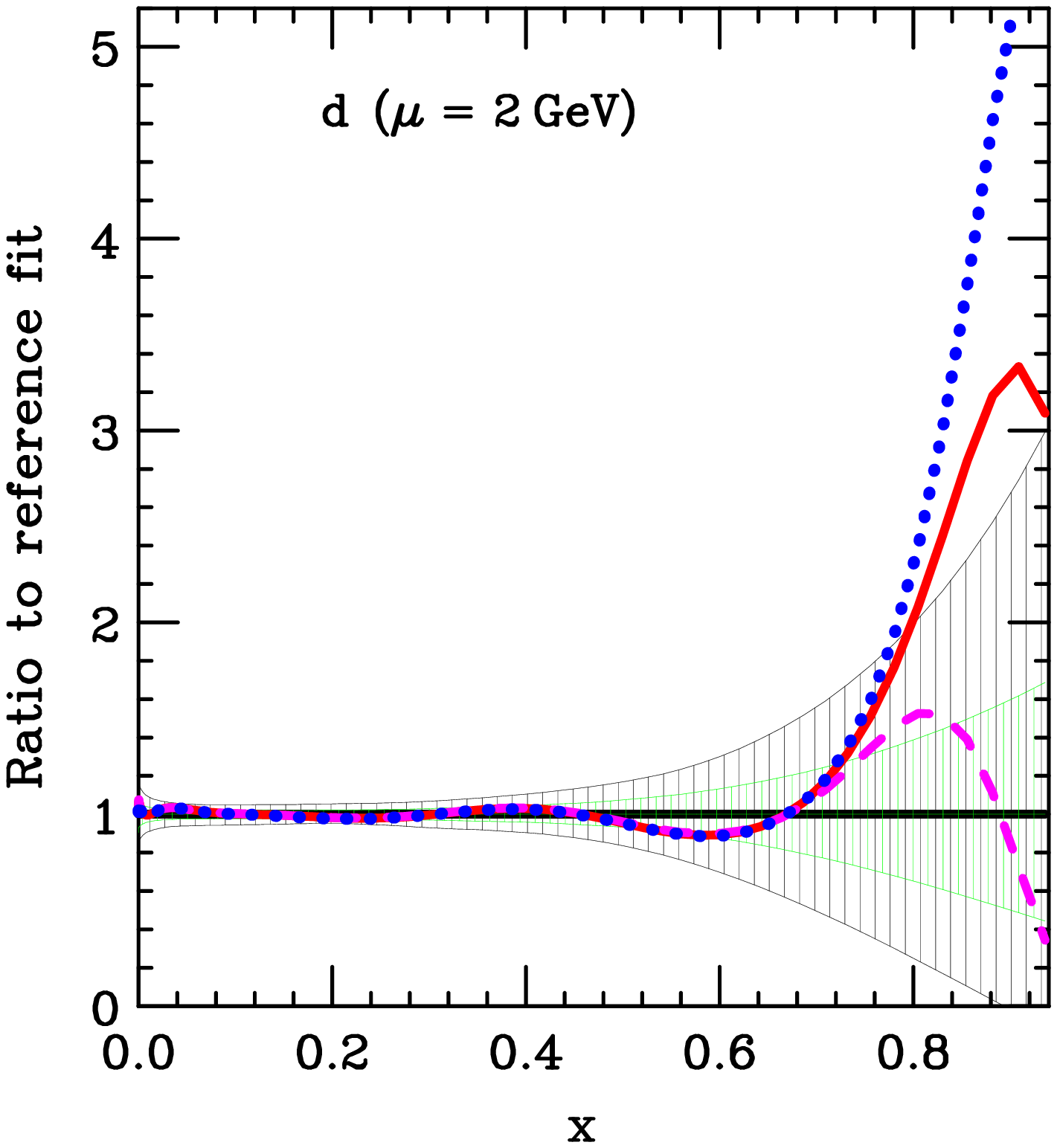}}
\hfill
 \resizebox*{0.21\textwidth}{!}{
\includegraphics[clip=true,scale=1.0]{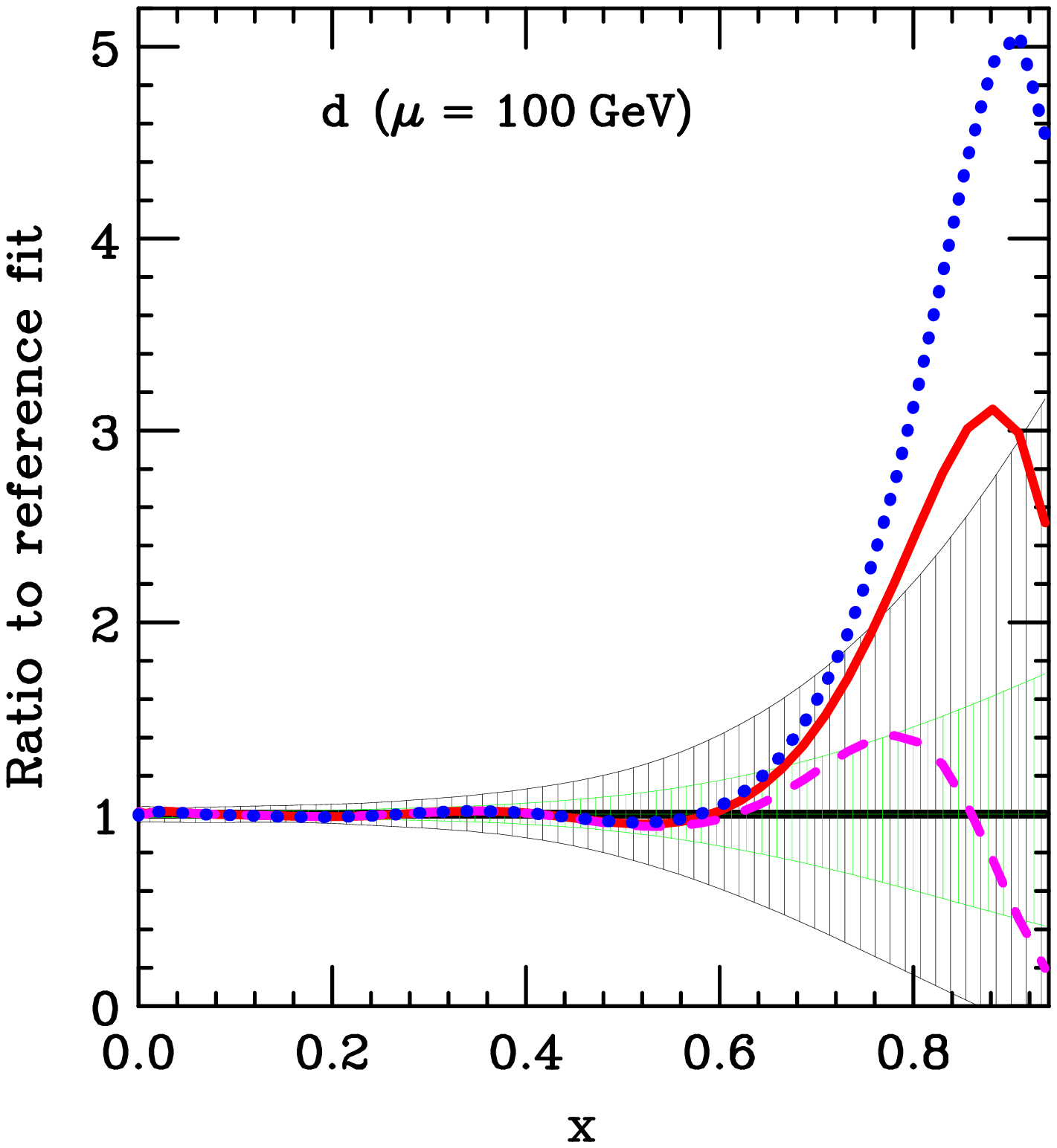}}
}
\mbox{
 \resizebox*{0.21\textwidth}{!}{
\includegraphics[clip=true,scale=1.0]{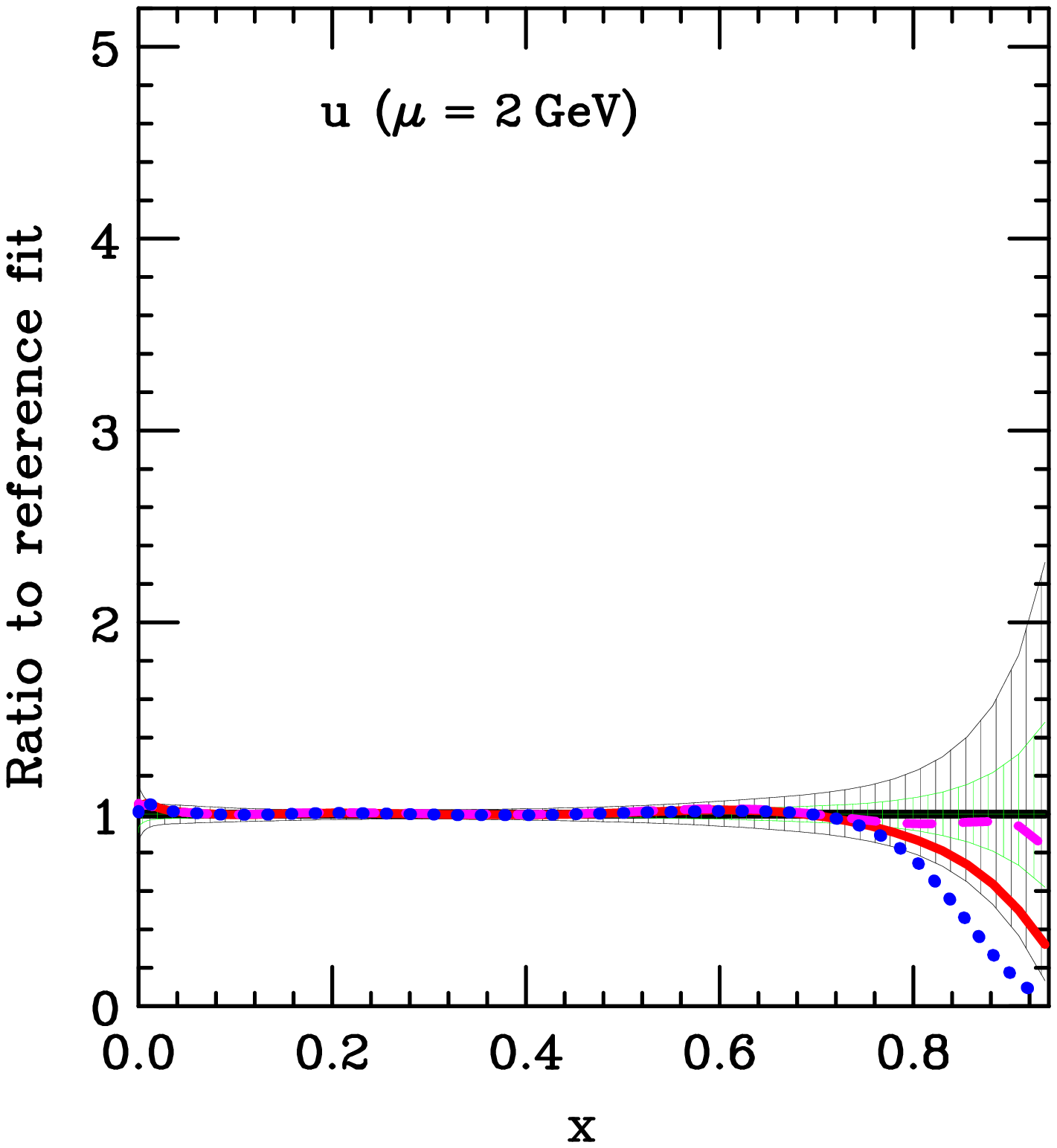}}
\hfill
 \resizebox*{0.21\textwidth}{!}{
\includegraphics[clip=true,scale=1.0]{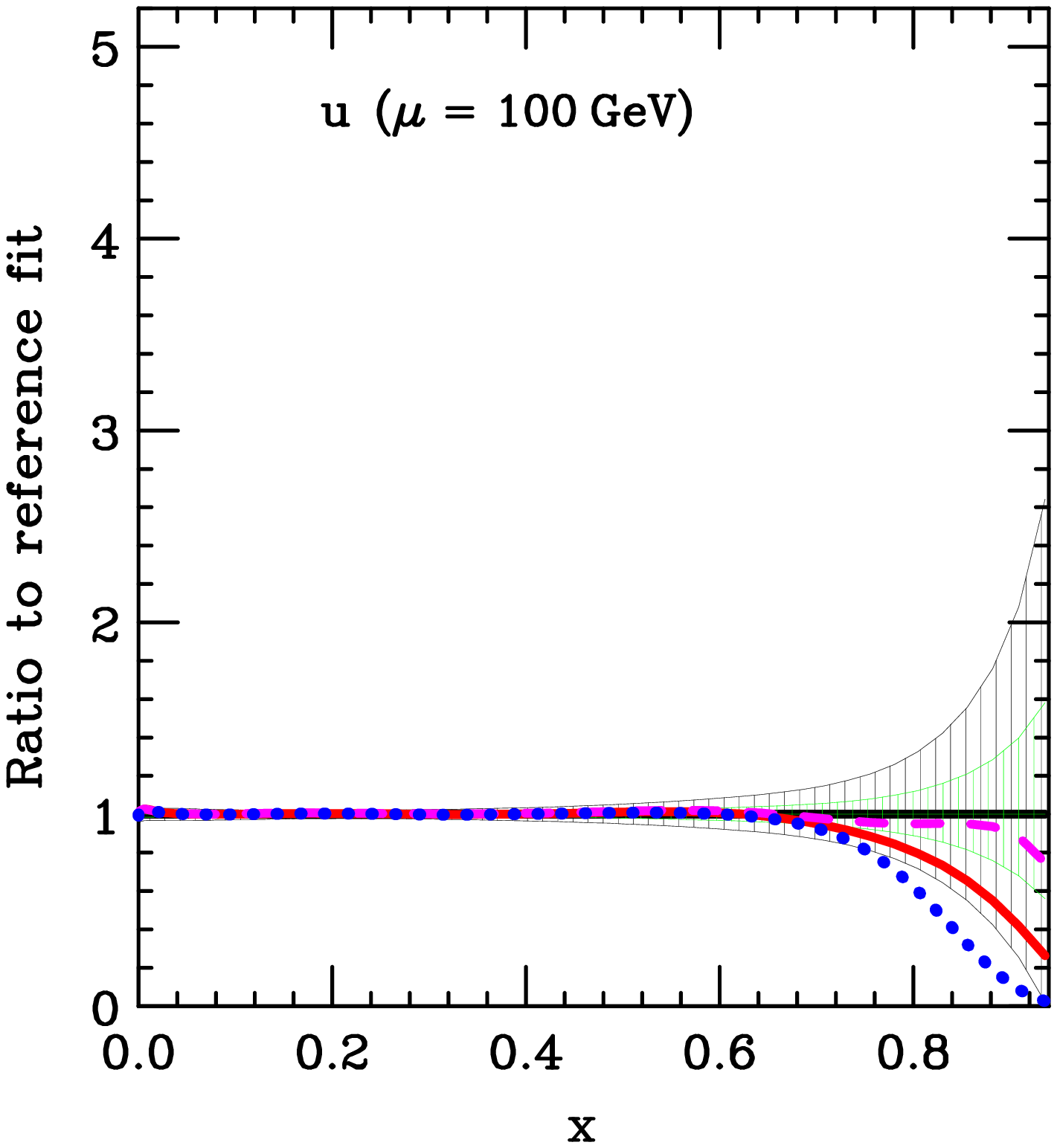}}
}
\vskip -10pt
 \caption{Like Fig.\ \ref{fig:figThree}, but displayed using 
linear scales to show the behavior at large $x$.
}
\vskip 10pt
 \label{fig:figFour}
\end{figure}

\begin{figure}[htb]
\vskip 20pt
\mbox{
 \resizebox*{0.24\textwidth}{!}{
\includegraphics[clip=true,scale=1.0]{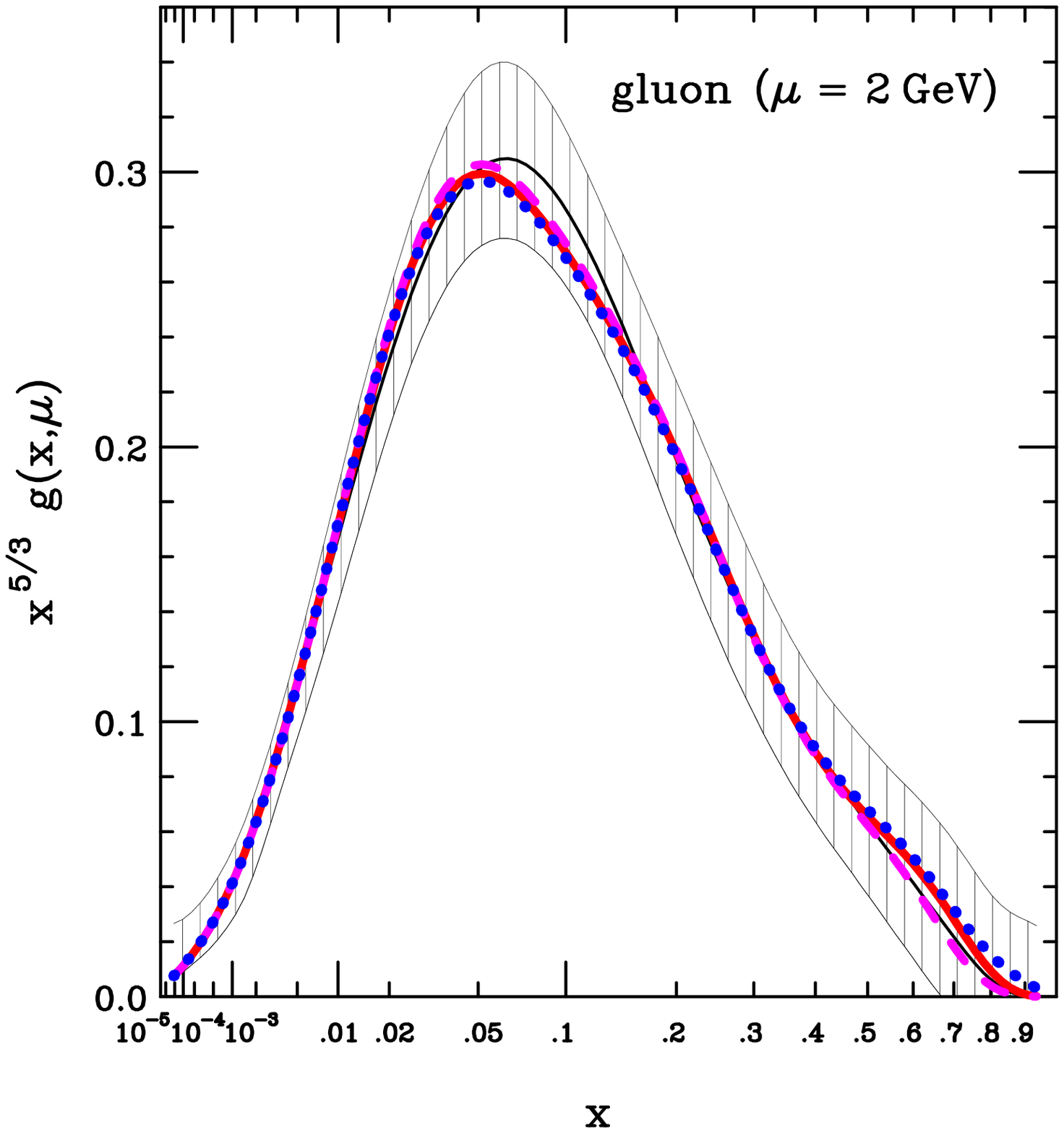}}
\hfill
 \resizebox*{0.24\textwidth}{!}{
\includegraphics[clip=true,scale=1.0]{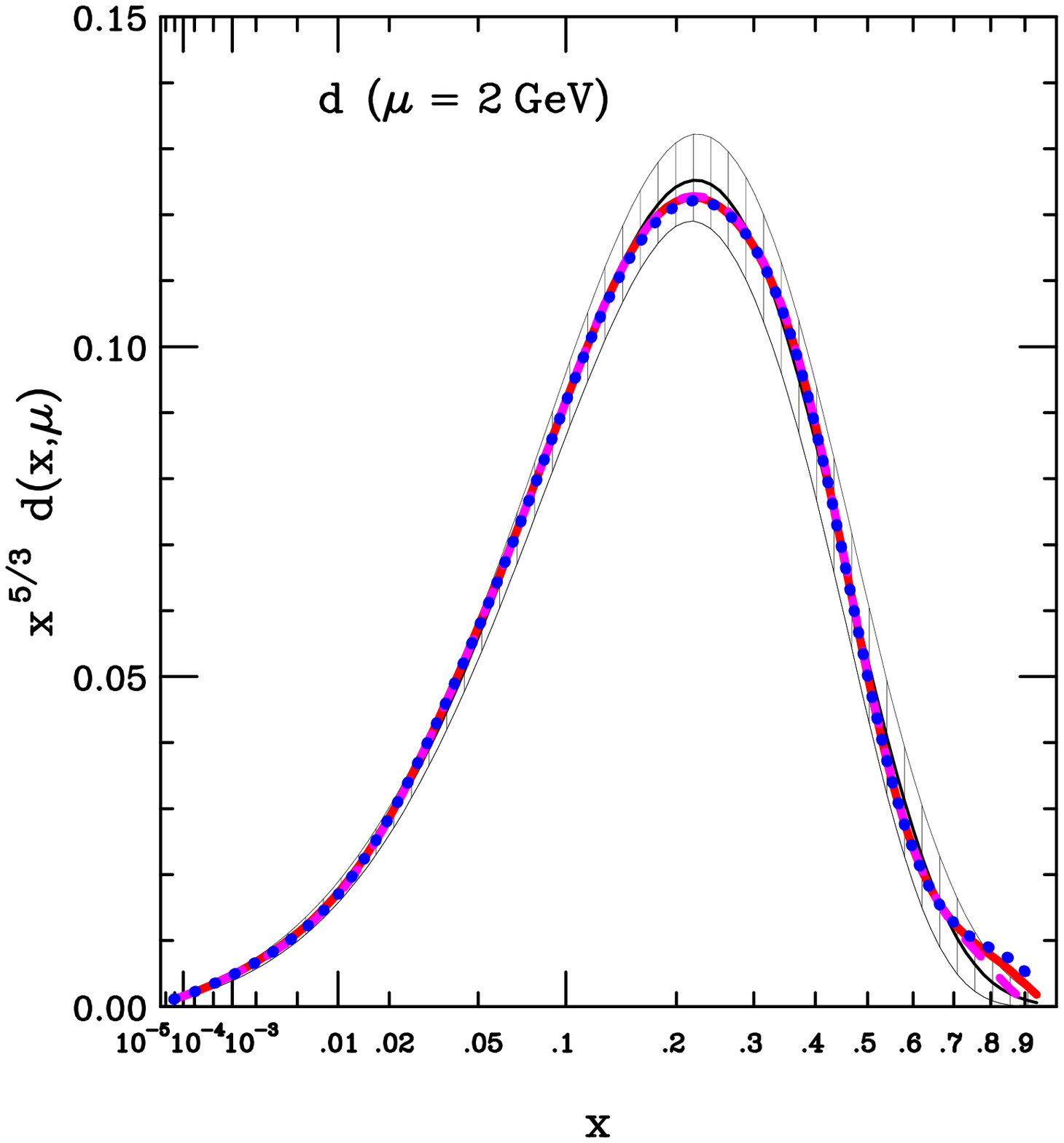}}
}
\mbox{
 \resizebox*{0.24\textwidth}{!}{
\includegraphics[clip=true,scale=1.0]{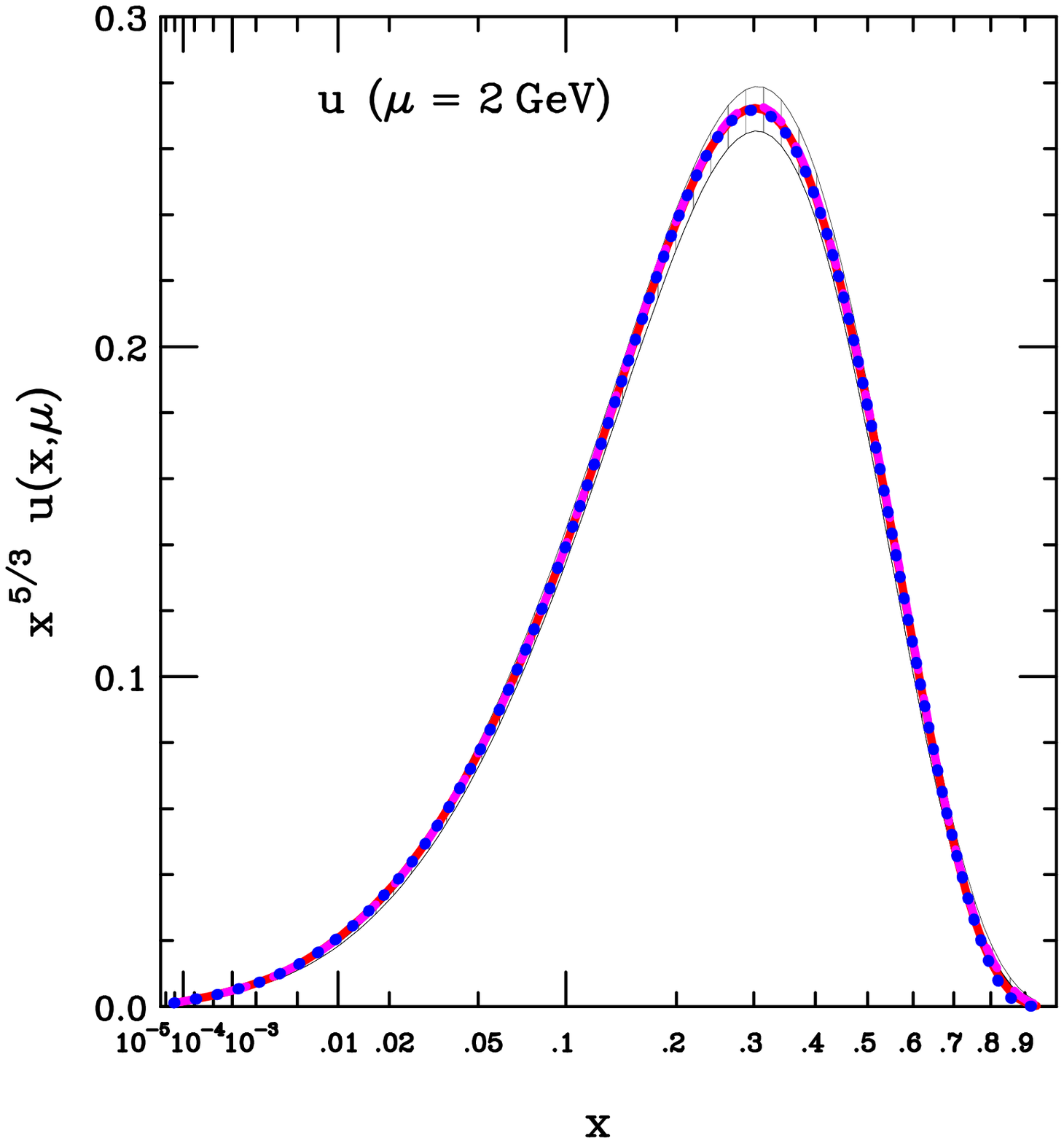}}
\hfill
}
\vskip -10pt
 \caption{Absolute gluon and quark distributions with 
uncertainties from CT10 (solid), and the three Chebyshev fits 
from Figs.\ \ref{fig:figThree} and \ref{fig:figFour}.
}
 \label{fig:figFive}
\vskip 10pt
\end{figure}

The wide shaded areas in Fig.\ \ref{fig:figThree} show the fractional 
uncertainty estimated at 90\% confidence in CT10 \cite{CT10}, 
which employs a $\Delta\chi_{\mathrm{eff}}^{\, 2} = 100$ criterion on a 
goodness-of-fit measure $\chi_{\mathrm{eff}}^{\, 2}$ that is defined as 
the sum of the usual $\chi^2$ plus supplemental ``penalty'' terms that 
are designed to force acceptable agreement to every data set over the 
entire allowed uncertainty range. 

The narrow shaded areas correspond to an otherwise similar fit, in which the 
uncertainty criterion is replaced by the pure $\Delta\chi^2 = 10$ condition
that was used in Sec.\ \ref{sec:fits}.
The ratio of these uncertainties is seen to be roughly a factor of 2.  It is 
not so large as the naive factor $\sqrt{100/10} = 3.16$, because of the penalty 
terms included in the goodness-of-fit measure for CT10, and because quadratic 
dependence of $\chi^2$ on the fitting parameters holds only rather close to the 
minimum in $\chi^2$.

The solid curve in each plot shows the result of a fit that was carried 
out in exactly the same way as the CT10 best fit, except for using the Chebyshev 
parametrization described in Appendix 2.  This fit has 84 free 
parameters---a few more than the Chebyshev fit of Sec.\ \ref{sec:fits}, because 
the simple CTEQ6.6 parametrization for strangeness was retained there.  It 
achieves a $\chi^2$ that is lower than CT10 by 105. \emph{This fit demonstrates 
that parametrization dependence introduces an uncertainty in CT10 that 
in some places approaches its $\Delta \chi^2 = 100$ uncertainty estimate. 
This result is consistent with the fact that the actual reduction in $\chi^2$ is 
close to that value.} This strong parametrization dependence can appear even 
in places where the fractional uncertainty is relatively small, such as in the 
$u$-quark distribution for $0.005 \lesssim x \lesssim 0.010$ at 
$\mu = 100 \, \mathrm{GeV}$. 

Figure \ref{fig:figFour} displays the same fits as Fig.\ \ref{fig:figThree}, 
using linear scales to reveal the behavior at large $x$.  The fractional 
uncertainty becomes large at very large $x$, because the available data 
provide little constraint there.
In spite of the very large uncertainty found by CT10 in that 
region, the actual uncertainty is still larger, as is seen in 
the case of the $d$-quark distribution for $x \gtrsim 0.8$.  This is not 
surprising, since the absence of experimental constraints 
implies that the behavior extracted at large $x$ is an extrapolation 
based mainly on the choice of parametrization.  The solid curve in this 
figure is a best fit using the Chebyshev method, while the dotted and 
dashed curves were obtained by adding a small penalty to $\chi^2$ in the 
Chebyshev method to push the 
gluon and $d$ quark distributions up or down relative to the 
$u$ quark distribution at large $x$.  These dotted and dashed fits have a 
$\chi^2$ that is only $5$ units higher than the Chebyshev best fit
(and hence lower than CT10 by $100$).  The very large 
difference between the dotted and dashed curves for the $d$-quark 
therefore corresponds to an uncertainty range of only $\Delta\chi^2 = 5$.  
The full $d$-quark uncertainty in the large-$x$ region must therefore be 
still much larger.

The flexibility of the Chebyshev parametrization is such that it could easily 
produce fits that contain unrealistically rapid variations in the PDFs as a 
function of $x$.  A necessary aspect of Chebyshev fitting, as described 
in Appendix 2, is therefore to include a penalty in the function that measures 
goodness-of-fit, to suppress any unwarranted fine structure.  
Figure \ref{fig:figFive} shows the same results as in 
Figs.\ \ref{fig:figThree} and \ref{fig:figFour} except that this 
time the absolute $g$, $d$, and $u$ PDFs are shown, instead of 
their ratio to the CT10 best fit.  
(The horizontal axis $x^{1/3}$ is used here to display both large and small 
$x$; while the weight factor $x^{5/3}$ included in the vertical axis makes 
the area under each curve proportional to its contribution to the momentum 
sum rule.)
This figure demonstrates that the method outlined in Appendix 2 to restrict 
the fits to functions that are reasonably smooth is successful.  
Further evidence of the smoothness of the Chebyshev fits can be seen in 
Figs.\ \ref{fig:figSix}--\ref{fig:figEight} of Sec.\ \ref{sec:LargeX}.

One might find the ``shoulder'' that appears in the central Chebyshev fit 
in Fig.\ \ref{fig:figFive} for the $d$-quark distribution 
at $x \gtrsim 0.8$---and perhaps the milder shoulder in $g(x)$ at 
$x \gtrsim 0.6$---to be unlikely features of nuclear 
structure.  These features are certainly not required by the data, since the 
dashed fits which do not have them have a larger $\chi^2$ by only 5 units;
but at present, there seems to be no strong theoretical basis to exclude them.

The Chebyshev best fit shown here has $\chi^2 = 2916$ for $2753$ data points.  
This is lower than the CT10 $\chi^2$ by $105\,$.  Since the 90\% confidence 
uncertainty in CT10 is estimated using a $\Delta \chi^2 = 100$ criterion 
(with modifications to require an acceptable fit to each individual data set), 
it is not surprising that the parametrization error we find for CT10 is 
comparable to the CT10 error estimate, except at extreme values of $x$.  
At very large $x$, where the PDFs are poorly determined, the uncertainty in 
CT10 needs to be expanded, as evidenced by Fig.\ \ref{fig:figFour}.

A large part of the decrease in $\chi^2$ produced by the Chebyshev fit comes 
from the 
BCDMS $\mu p \to \mu X$ \cite{Benvenuti:1989rh} ($-21$) and 
BCDMS $\mu d \to \mu X$ \cite{Benvenuti:1989fm} ($-16$) experiments, 
which are particularly sensitive to $u$ and $d$ quark distributions at 
large $x$.  Other important improvements in the fit are to the combined 
HERA-1 data set \cite{HERAcombined} ($-17$), to the 
CDF run 2 inclusive jet measurement \cite{CDFjet} ($-15$), 
and to the E866 Drell-Yan pp data \cite{Webb:2003ps} ($-11$).

\section{PDF behavior at large x 
\label{sec:LargeX}}

\begin{figure}[htb]
\vskip 20pt
\mbox{
 \resizebox*{0.22\textwidth}{!}{
\includegraphics[clip=true,scale=1.0]{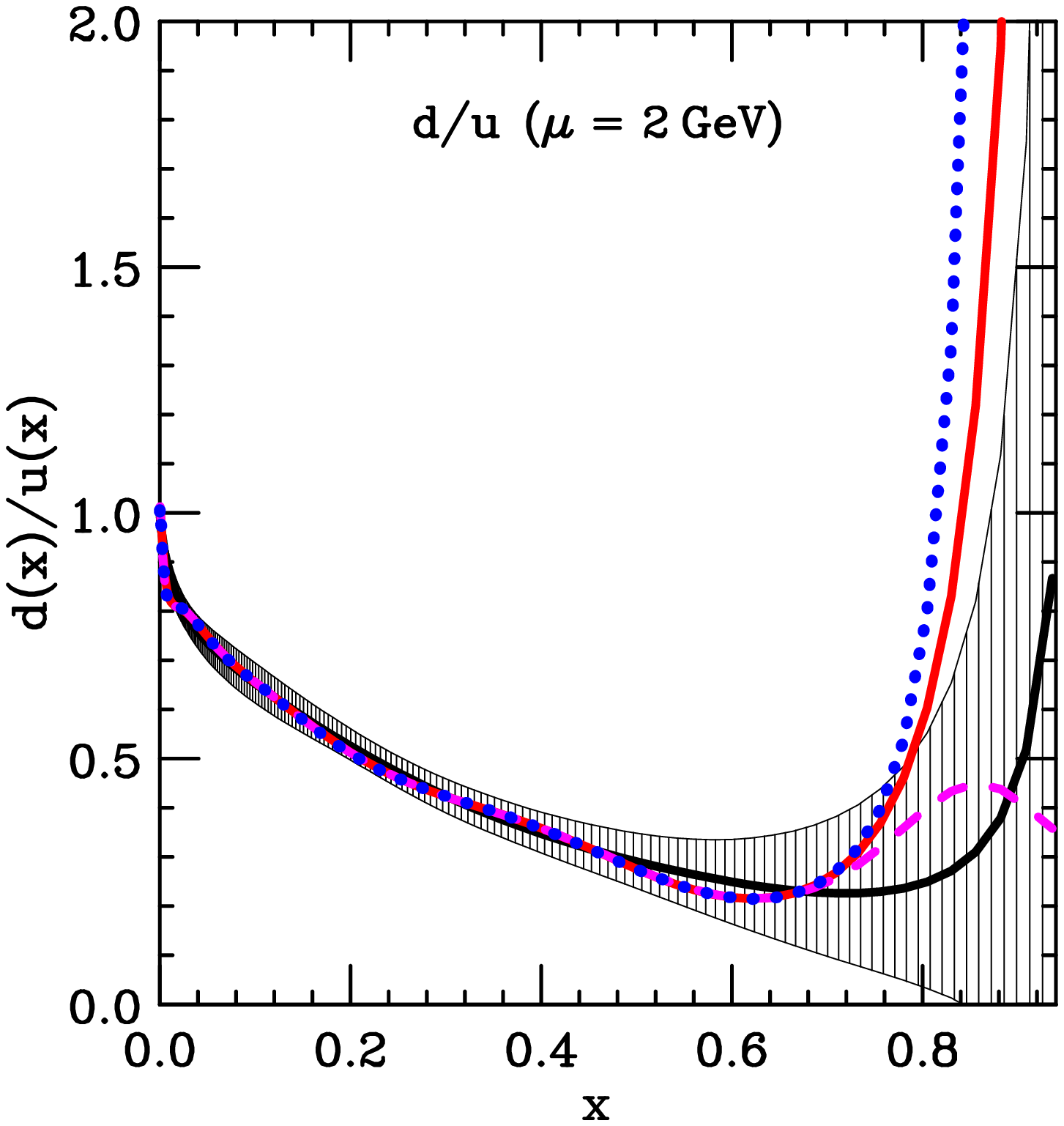}}
\hfill
 \resizebox*{0.22\textwidth}{!}{
\includegraphics[clip=true,scale=1.0]{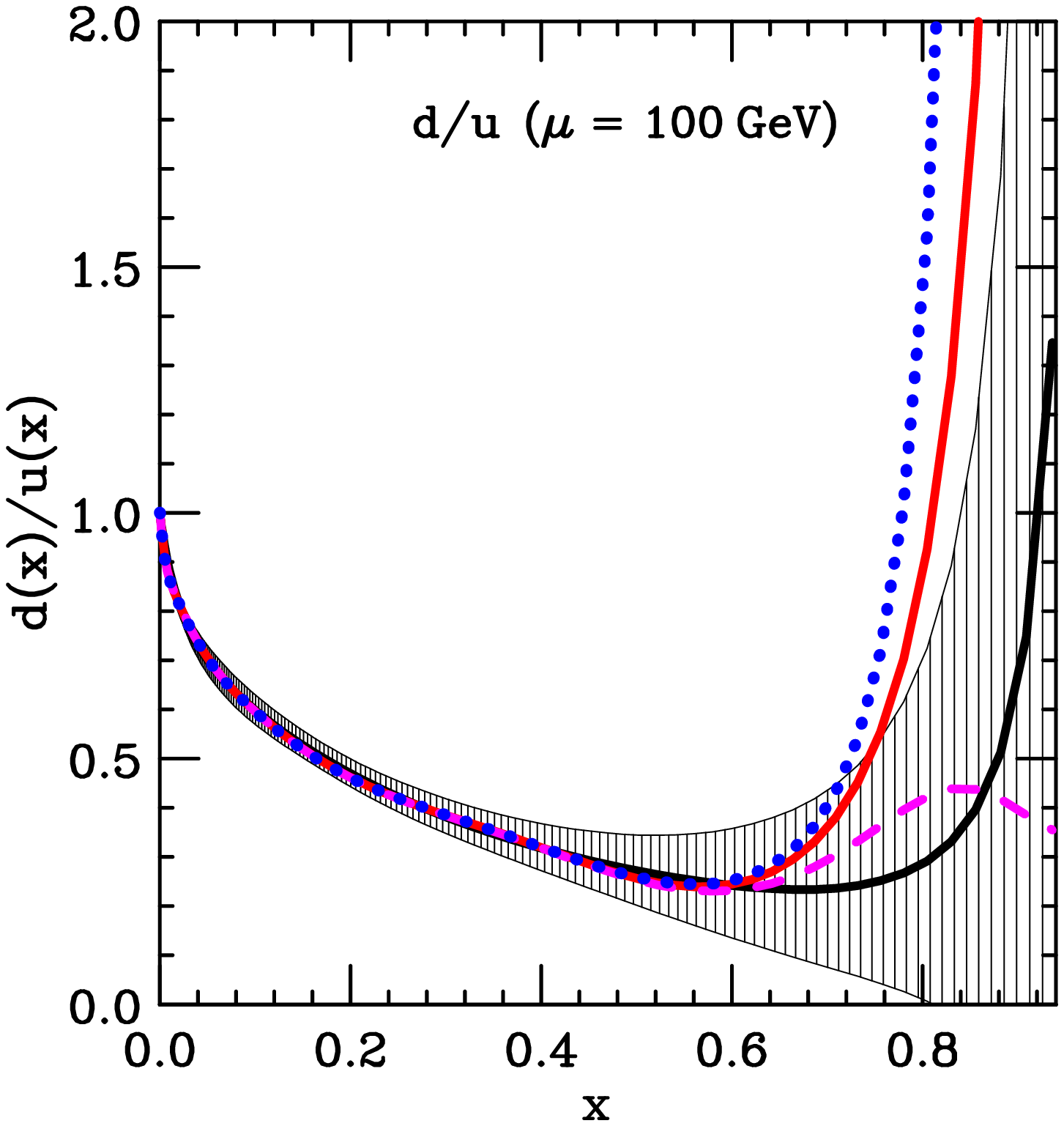}}
}
\vskip -10pt
 \caption{Quark ratio $d(x)/u(x)$ at scales
$\mu = 2 \, \mathrm{GeV}$ and $\mu = 100 \, \mathrm{GeV}$.
Shaded area is CT10 with uncertainty; curves are the same 
Chebyshev fits shown in Figs.\ \ref{fig:figThree}--\ref{fig:figFive}.
}
 \label{fig:figSix}
\vskip 10pt
\end{figure}

The parton distributions at large $x$ are not well 
constrained by data.  For example, Fig.\ \ref{fig:figSix} shows a 
large uncertainty of the $d/u$ ratio at $x \gtrsim 0.8$.  Recall that 
the dotted and dashed extreme curves shown here represent an increase 
in $\chi^2$ by only 5 units above the Chebyshev best fit, so the full 
uncertainty must be considerably larger than what is spanned by those 
curves.

\begin{figure}[htb]
\vskip 20pt
\mbox{
 \resizebox*{0.23\textwidth}{!}{
\includegraphics[clip=false,scale=1.0]{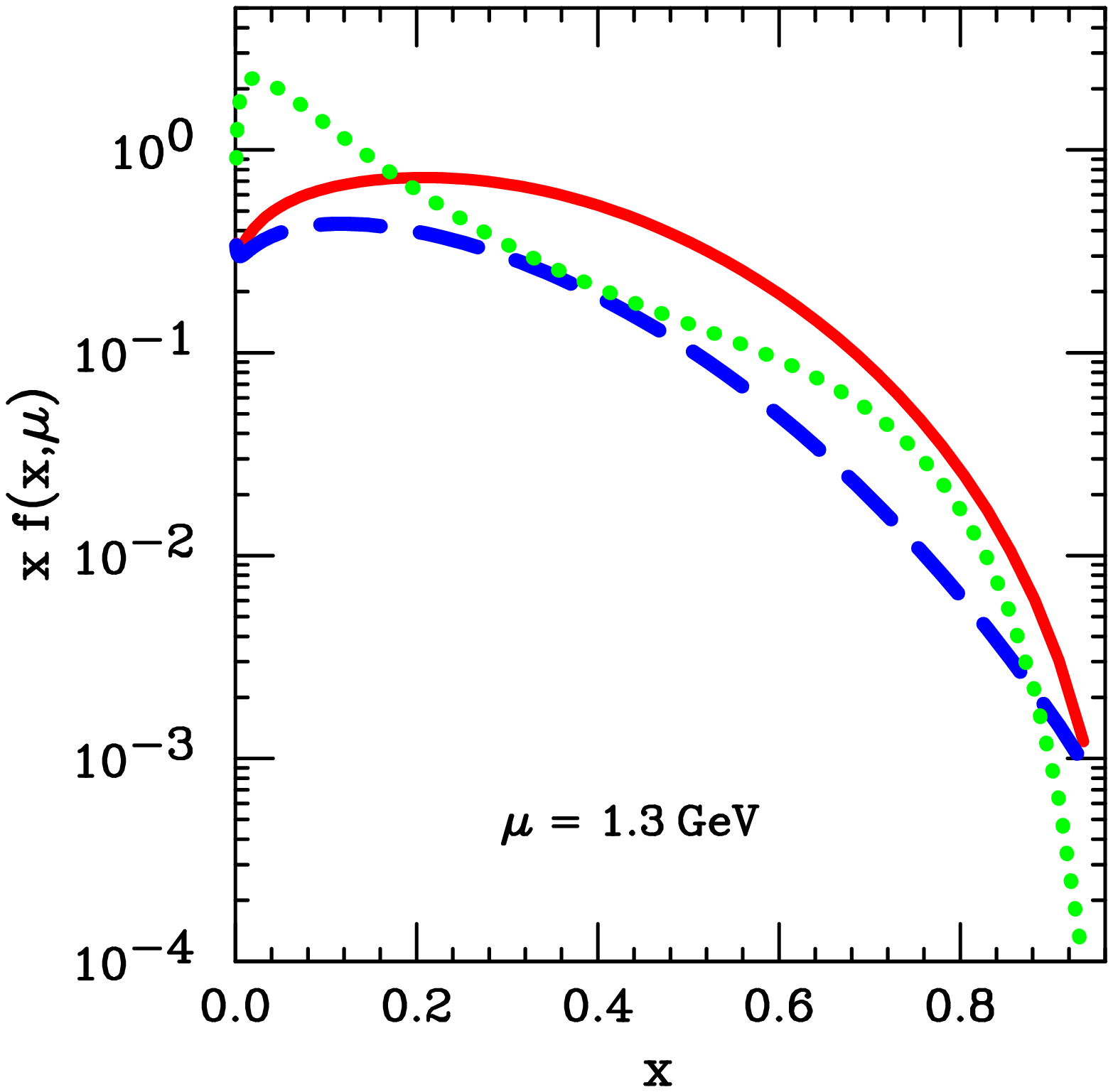}}
\hfill
 \resizebox*{0.23\textwidth}{!}{
\includegraphics[clip=false,scale=1.0]{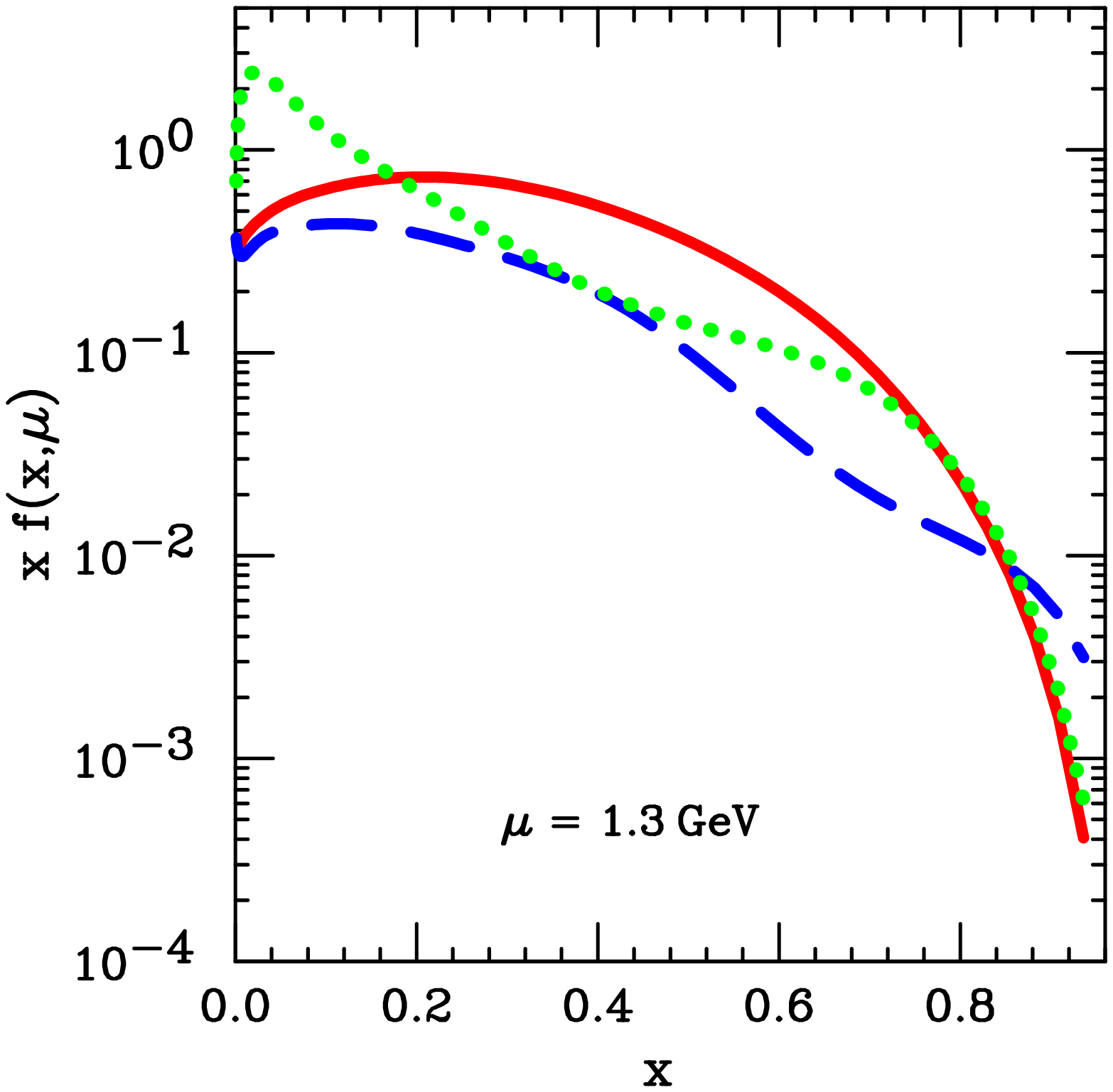}}
}
\mbox{
 \resizebox*{0.23\textwidth}{!}{
\includegraphics[clip=false,scale=1.0]{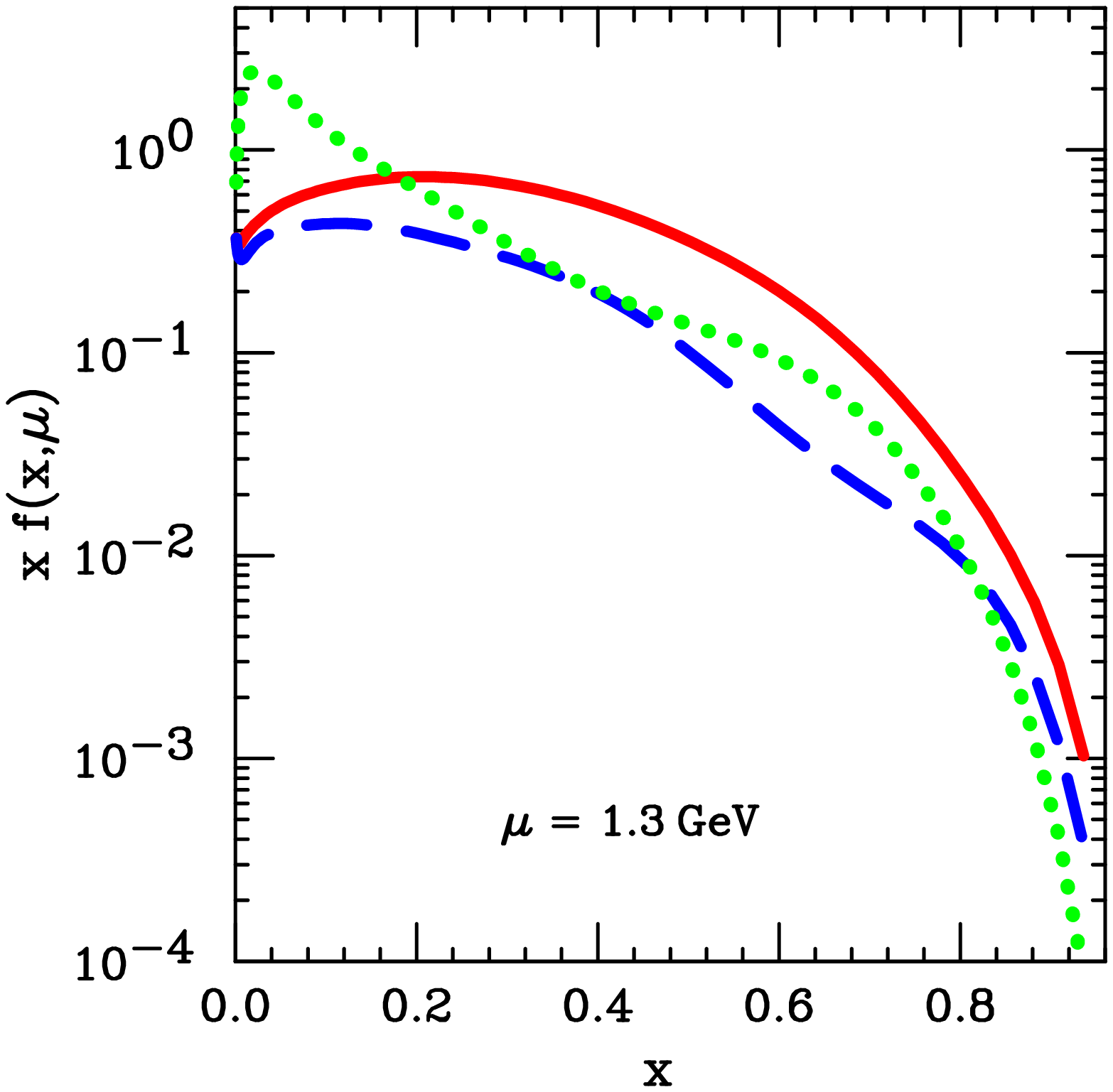}}
\hfill
 \resizebox*{0.23\textwidth}{!}{
\includegraphics[clip=false,scale=1.0]{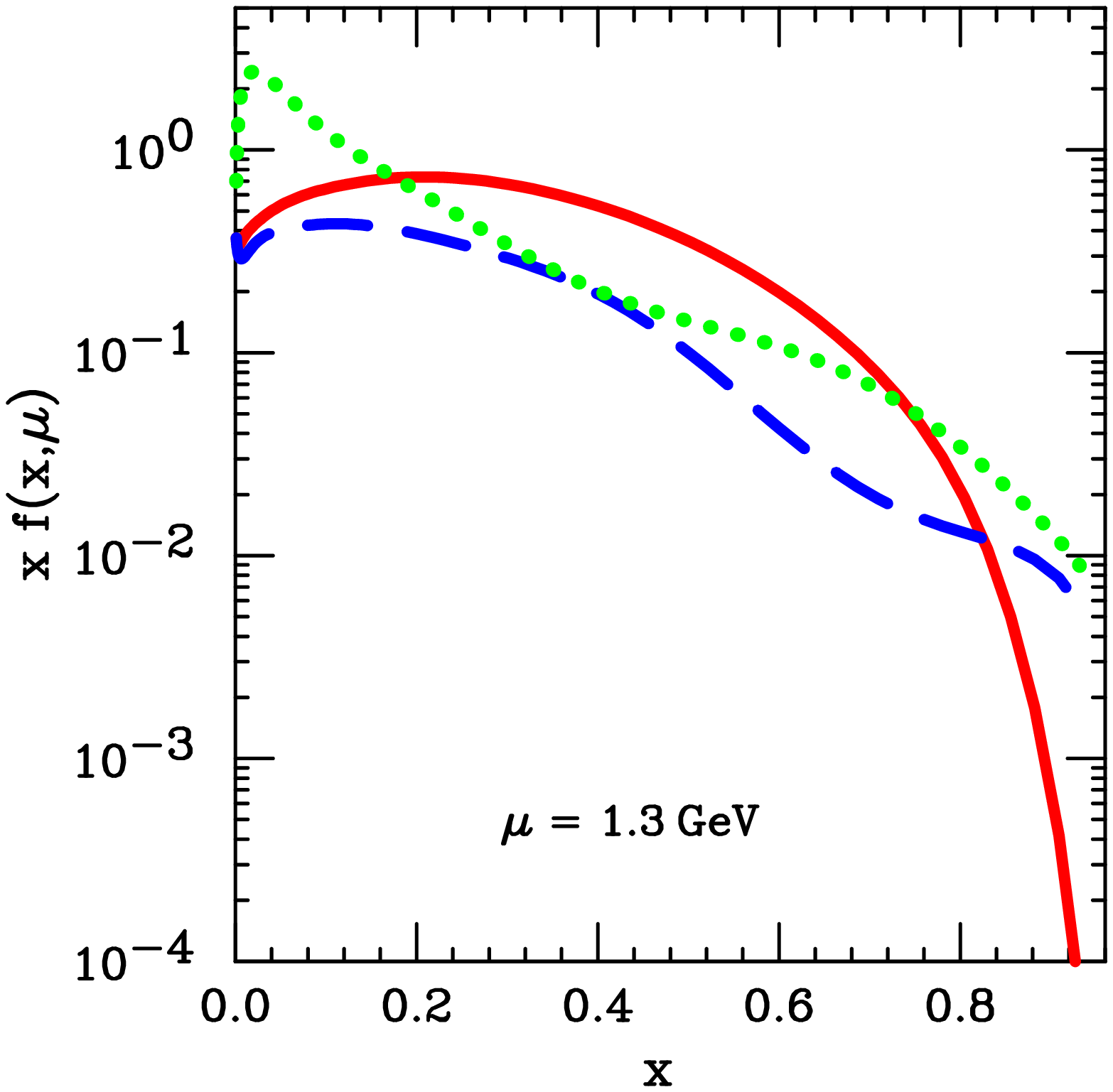}}
}
\vskip -10pt
 \caption{Up quark (solid), down quark (dashed), and gluon (dotted) 
distributions at $\mu = 1.3 \, \mathrm{GeV}$  
from CT10 (upper left), and the three Chebyshev fits, which 
have $\chi^2$ lower than CT10 by $105$, $100$, and $100$.
}
 \label{fig:figSeven}
\vskip 10pt
\end{figure}

\begin{figure}[htb]
\vskip 20pt
\mbox{
 \resizebox*{0.23\textwidth}{!}{
\includegraphics[clip=true,scale=1.0]{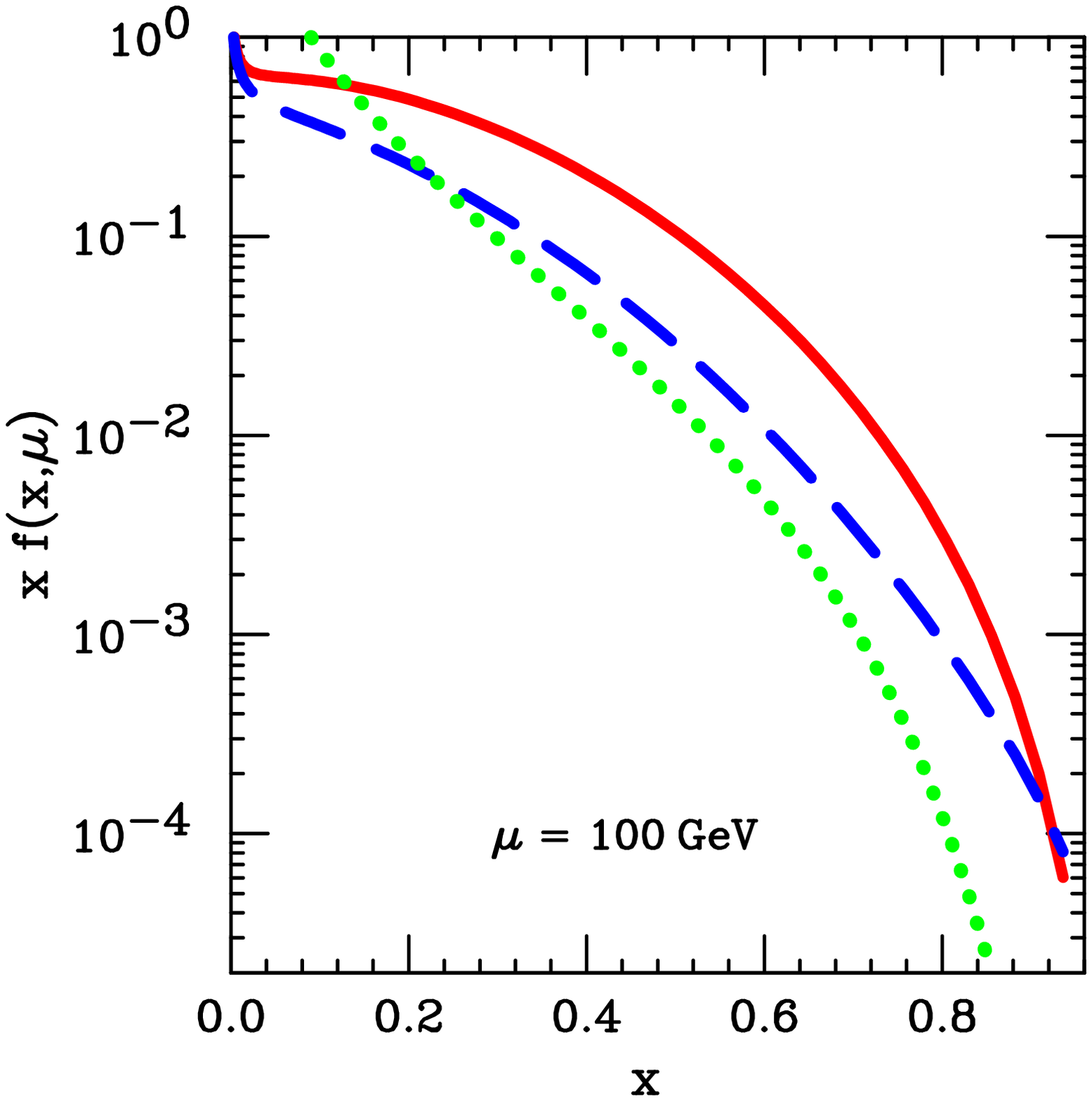}}
\hfill
 \resizebox*{0.23\textwidth}{!}{
\includegraphics[clip=true,scale=1.0]{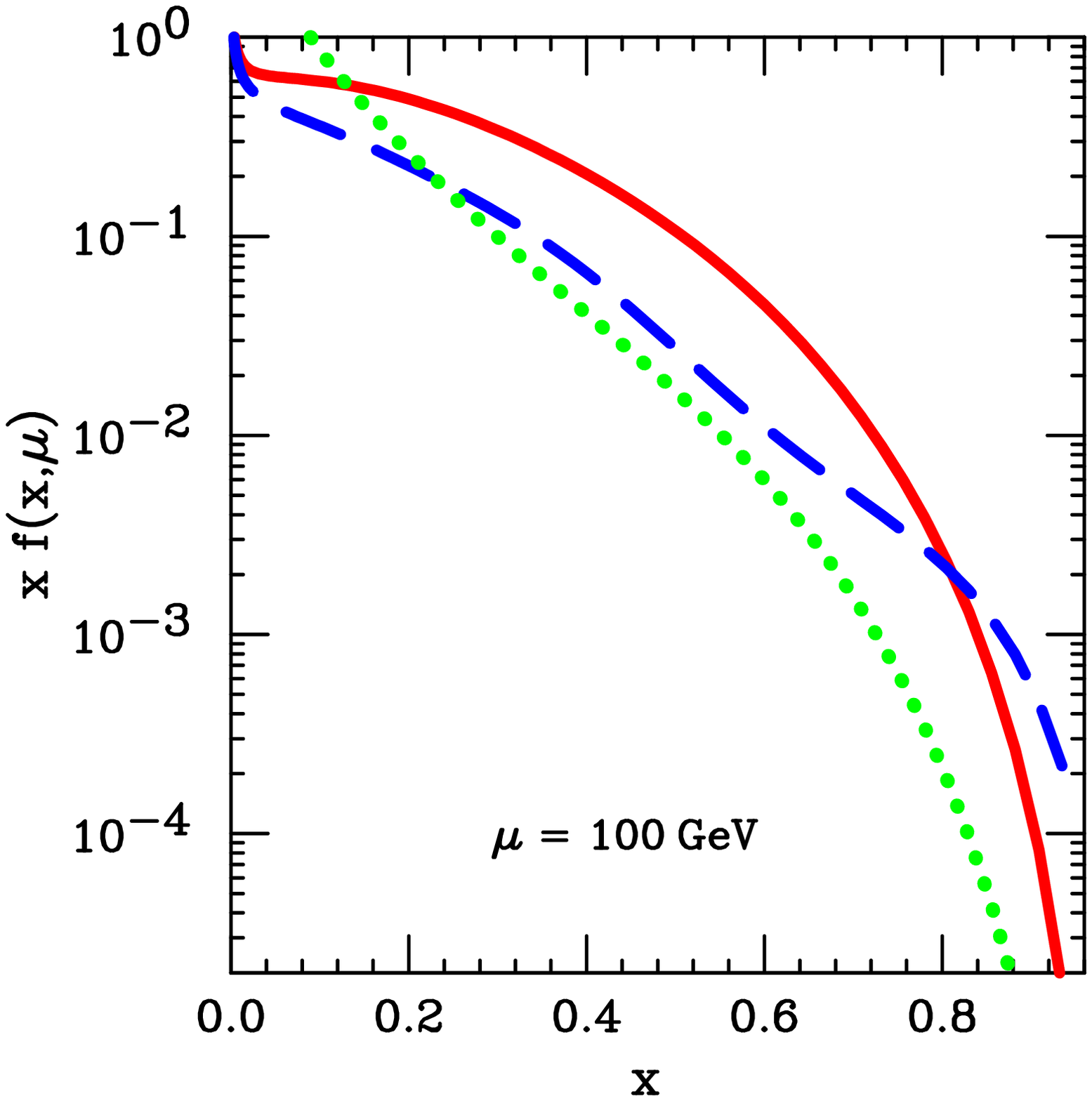}}
}
\mbox{
 \resizebox*{0.23\textwidth}{!}{
\includegraphics[clip=true,scale=1.0]{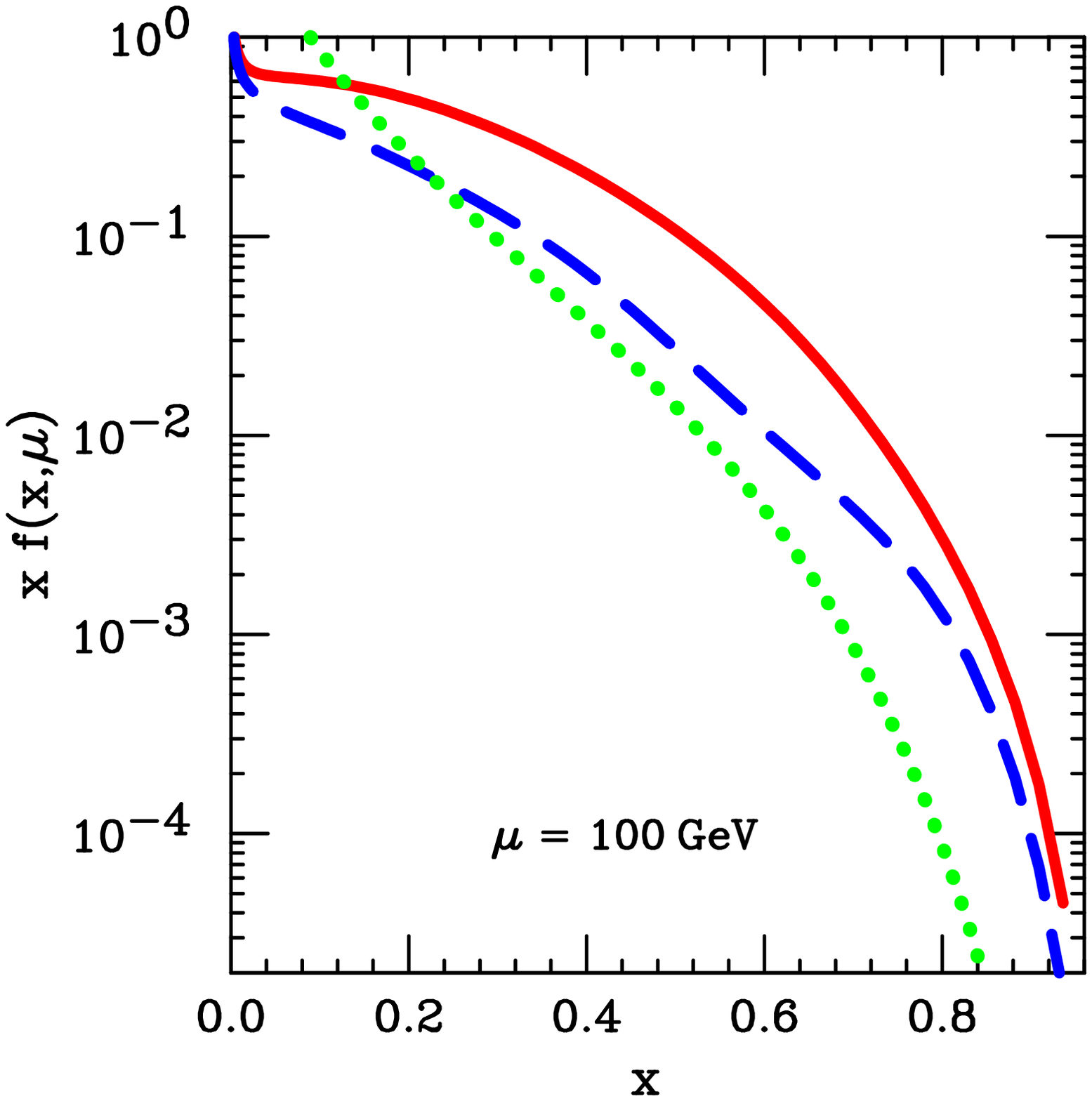}}
\hfill
 \resizebox*{0.23\textwidth}{!}{
\includegraphics[clip=true,scale=1.0]{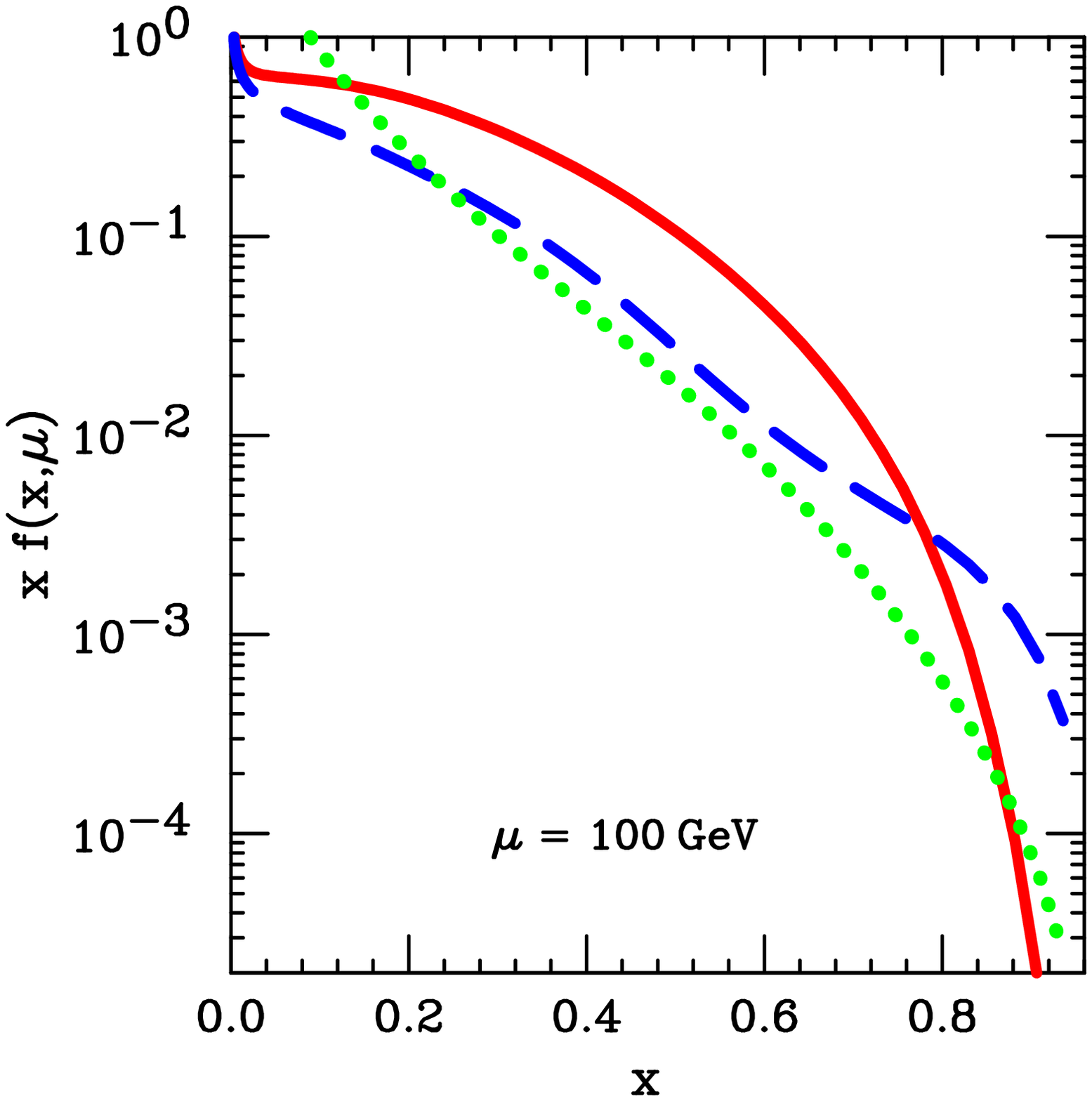}}
}
\vskip -10pt
 \caption{Like Fig.\ \ref{fig:figSeven}, but at scale 
$\mu = 100 \, \mathrm{GeV}$.
}
 \label{fig:figEight}
\vskip 10pt
\end{figure}

The large-$x$ behavior of the best fits for $u$, $d$, and 
$g$ at $\mu = 1.3 \, \mathrm{GeV}$ are compared directly in
the first panel of Fig.\ \ref{fig:figSeven} for CT10, and in 
the other three panels of Fig.\ \ref{fig:figSeven} for the 
three Chebyshev fits that 
were shown in Figs.\ \ref{fig:figThree}--\ref{fig:figSix}.
We see that the intuitively expected order $u(x) \gg d(x) \gg g(x)$ 
in the limit $x \to 1$ is consistent with the fitting, since it 
appears in third panel of Fig.\ \ref{fig:figSeven}; but that this 
behavior is not 
required by it---\textit{e.g.}, 
$d(x) \gg u(x) \approx g(x)$ in the second panel, and 
$g(x) \approx d(x) \gg u(x)$ in the fourth panel.
\emph{It is possible that theoretical ideas from nonperturbative physics such 
as $u(x) \gg d(x) \gg g(x)$ should be imposed to reduce the uncertainty of 
the fitting, by including a ``soft constraint,''} which would be implemented 
by adding an appropriately defined penalty to the effective $\chi^2$.  
However, all of these fits share the feature that $g(x)$ is comparable to or 
larger than $d(x)$ in the rather broad range $0.3 \lesssim x \lesssim 0.8\,$.
\emph{That valence-like behavior of the gluon poses an interesting challenge 
to be explained by hadron structure physics; but meanwhile, it demonstrates that 
naive expectations about the parton content of the proton can be unreliable.}
Figure \ref{fig:figEight} shows that the ``natural'' large-$x$ order 
$u(x) > d(x) > g(x)$ is restored by DGLAP evolution for factorization scales
above $\mu = 100 \, \mathrm{GeV}$, except possibly at extremely large $x$.

\section{Conclusion 
\label{sec:conclusion}}

Increasing the flexibility of the input parametrizations used in PDF analysis,
using a method based on Chebyshev polynomials, has revealed a significant 
source of uncertainty in previous PDF determinations caused by parametrization 
dependence.  As demonstrated in Figs.\ \ref{fig:figThree} and \ref{fig:figFive}, 
the parametrization error is in many places as large as the uncertainty 
estimated, via $\Delta\chi^2 = 10$, on 
the basis of conflicts observed between the different data sets used 
in the analysis \cite{Consistency}.
At very large $x$, where the fractional PDF uncertainty is large, 
the parametrization error even becomes large compared to the modified 
$\Delta\chi^2 = 100$ uncertainty estimate used in CT10 \cite{CT10}, 
which was intended to represent a 90\% confidence interval.
These parametrization effects persist 
up to scales of $\mu = 100 \, \mathrm{GeV}$ and beyond, so they are 
significant for predictions of important background and discovery processes 
at the Tevatron and LHC.

The hypothetical example shown in the left panel of 
Fig.\ \ref{fig:figOne} raised the spectre 
of large possible shifts in a PDF fit when a new degree of freedom is 
included in the parametrization.  As stated earlier, that degree of freedom 
must generally correspond to a linear combination of 
\textit{several} new fitting parameters, since the reduction in $\chi^2$ 
depends on modifying more than one flavor in some particular correlated manner.  
We have seen that enhancing parametrization freedom by the Chebyshev method of 
Appendix 2 reduces $\chi^2$ for the fit to the CT10 data set by 105.  
That decrease is nearly equal to the tolerance criterion used to estimate the 
uncertainty in CT10, so it is not surprising that the corresponding changes 
in the PDFs can be comparable to the CT10 uncertainty estimate. 
\emph{This result 
lends support to the use of $\Delta\chi^2 \sim 100$ in CT10, while suggesting 
the possibility that a somewhat tighter criterion could be used, once the 
Chebyshev-style parametrizations have been incorporated into the uncertainty 
analysis.}

The right panel of Fig.\ \ref{fig:figOne} illustrates how large changes in 
the results could 
arise in principle even from a very modest decrease in $\chi^2$.  
Figure \ref{fig:figFour} shows that this can actually happen in practice, 
in regions where the results of fitting are dominated by parametrization 
assumptions because the data provide little constraint. 
In particular, 
the solid (red) curves in Figs.\ \ref{fig:figFour}--\ref{fig:figSix} show a 
best fit obtained using the Chebyshev method; while the dotted and dashed 
curves, which have rather different behavior at large $x$, are fits made 
by including mild constraints on the large-$x$ behavior.  Those constraints 
are so mild that they increase the overall $\chi^2$ by only 5 units.

The parametrization effects discussed in this paper come from
increasing the flexibility of the functional forms used to approximate
the PDFs at the chosen starting scale $\mu_0$ for their evolution,
without altering the various discrete assumptions that went into choosing
those forms.  For example, a much wider uncertainty range would be 
permitted for $\bar{d}(x)/\bar{u}(x)$ in Fig.\ \ref{fig:figSix} if 
the assumption $r = 1$ were relaxed, where
$r = {\displaystyle \lim_{x \to 0}} (\bar{d}(x,\mu_0)/\bar{u}(x,\mu_0))$.
Indeed, the range $0.4 < r < 1.2$ would be permitted by an increase in 
$\chi^2$ as small as $\Delta\chi^2 = 5$.  (The parameter $r$ is treated 
as a free parameter in the Alekhin2002 PDFs \cite{Alekhin}: its value in the 
central fit is approximately $0.8\,$.)
As a still more blatant example of these choices, all of the fits discussed 
here have $s(x,\mu) = \bar{s}(x,\mu)$ as a result of a simplifying assumption 
in the parametrization. Dropping that approximation would allow a very wide 
range of the asymmetry $(s(x) - \bar{s}(x))/(s(x) + \bar{s}(x))$.

In future analyses, it will be important to properly combine the 
parametrization uncertainty with the other sources of uncertainty that have 
previously been included in the Hessian method.  A step in that direction 
could be made by using a very flexible parametrization such as the Chebyshev 
one to determine the best fit, but then to freeze enough of the parameters 
to make the usual Hessian method tractable.  This would be an extension of 
the approach already used in MSTW \cite{MSTW08} and CTEQ \cite{Hessian} fits, 
wherein one or two parameters for each flavor are frozen at their best-fit 
values before the Hessian eigenvector method is carried out.  Once the 
parametrization error has been reduced by means of more flexible 
parametrizations, it should become possible to apply a tighter uncertainty 
criterion, \textit{e.g.}, comparable to the $\Delta\chi^2 \approx 10$ for 
90\% confidence that is suggested by the observed level of 
consistency \cite{Consistency} among input data sets.

An alternative method to avoid parametrization dependence is offered by the 
NNPDF approach \cite{NNPDF}, in which the PDFs at $\mu_0$ are represented 
using a neural network model that contains a very large number of effective 
parameters.  Broadly speaking, the uncertainties estimated in this way appear 
to be consistent with the results presented here.  A more detailed comparison 
will require allowing for differences in assumptions about the nonperturbative 
hadronic physics, such as positivity of the input distributions.
Attention will also have to be paid to the choice of assumptions about behaviors 
in the $x \to 0$ and $x \to 1$ limits, which are imposed in the NNPDF approach 
by ``preprocessing exponents.'' This will be undertaken in a future work.

\begin{acknowledgments}
I thank my TEA (Tung \textit{et al.})
colleagues J.~Huston, H.~L.~Lai, P.~M.~Nadolsky, and C.--P.~Yuan
for discussions of these issues.  I thank Stefano Forte and Robert Thorne
for helpful correspondence and discussions.
This research was supported by National Science Foundation grants PHY-0354838 
and PHY-08555561. 
\end{acknowledgments}

\section*{Appendix 1: General form of $\chi^2$ near its minimum}
This Appendix shows that the qualitative behavior of $\chi^2(y,z)$ hypothesized 
in Fig.\ \ref{fig:figOne} arises under rather general assumptions.

Let us assume as usual that $\chi^2$ can be approximated 
in the neighborhood of its minimum by Taylor series through second order.  
To find the uncertainty of a particular variable, we can assume by means of a 
linear transformation of that variable, that we are interested in the value of 
a parameter $z$ for which $\chi^2 = z^2 \, + \, C$ at $y=0$, as in 
Eq.~(\ref{eq:eq1}).  
Now let $y$ represent an additional fitting parameter that was previously held 
fixed at $0$.  
By Taylor series, the expression for $\chi^2$ expands to become
\begin{equation}
\chi^2 \, = \, z^2 \, + \, y^2 \, + \, 2 A z y \, + \, 2 B y \, + \, C \, ,
\label{eq:eq2}
\end{equation}
where the coefficient of $y^2$ was chosen to be $1$ without loss of generality,
by scaling that variable.  Equation (\ref{eq:eq2}) implies that the contours of 
constant $\chi^2$ are ellipses whose major and minor axes make an angle 
of $\pm 45^\circ$ with respect to the $y$ and $z$ axes, as in the specific 
examples of Fig.~\ref{fig:figOne}.  The ratio of minor axis to major axis of 
the ellipse is $\sqrt{(1 - |A|)/(1 + |A|)}\,$, and $|A| < 1$ is required, since 
$\chi^2$ must have a minimum.  The minimum of $\chi^2$ occurs at 
\begin{equation}
     z_0 = A B / (1 - A^2) \, , \quad 
     y_0 = - B / (1 - A^2) \; ,
\end{equation}
and its value there is 
\begin{equation}
    \chi_0^{\, 2} \, = \, C \, - \, D \, , \; \mbox{where} \;
    D \, = \, B^2 / (1 - A^2) \; .
\label{eq:eq7}
\end{equation}
Relative to the $y=0$ situation given by  Eq.~(\ref{eq:eq1}), introducing the 
additional parameter $y$ thus allows the best-fit $\chi^2$ to be lowered by $D$. 
At the same time, it demands that the uncertainty range for $z$ be extended at 
least far enough to include $z_0$, and hence demands 
$\Delta\chi^2 > z_0^{\, 2}\,$ in the $y=0$ model. 

The hypothetical examples shown in Fig.\ \ref{fig:figOne} correspond to 
$A = B/5 = - \, \sqrt{2/3}\,$  $\, \Rightarrow \,$ $(z_0 = 10$, $D = 50$), 
and $A = 2 B = - \, \sqrt{20/21}\,$ $\, \Rightarrow \,$ ($z_0 = 10$, $D = 5$).  

\section*{Appendix 2: Details of a specific Chebyshev method}
This Appendix describes a specific method that uses Chebyshev polynomials 
to parametrize the parton distributions at starting scale $\mu_0$,
in a manner that allows great freedom in the functions, while 
maintaining their expected smoothness.

Each flavor is parametrized by the form (\ref{eq:fofx}),
with $p(x)$ given by (\ref{eq:tseries}) with $y = 1 - 2 \sqrt{x}$.  
For the studies presented this paper, $n=12$ was used for 
each flavor.  There are 6 flavors to be parametrized 
($g$, $u_v$, $d_v$, $\bar{u}$, $\bar{d}$, and $\bar{s}$, with 
$s(x) = \bar{s}(x)$ assumed), so this leads to 72 fitting parameters.  

To facilitate studying the effect of imposing constraints such as  
$\bar{d}(x)/\bar{u}(x) \to 1$ at $x \to 0$, instead of parametrizing 
$\bar{u}(x)$ and $\bar{d}(x)$ separately, 
the sum $\bar{u}(x) + \bar{d}(x)$ was parametrized in the 
manner described above for the other flavors, while the ratio 
$\bar{d}(x)/\bar{u}(x)$ was parametrized by 
\begin{eqnarray}
\bar{d}(x)/\bar{u}(x) \,  = \, 
\exp\left(c_0 \, + \, \sum_{j=1}^{13} c_j \, (T_j(y) - 1)\, \right) \; .
\label{eq:dbaroubar}
\end{eqnarray}
This method provides freedom for $\bar{u}(x)$ and $\bar{d}(x)$ separately 
that is comparable to the great freedom for other flavors, while allowing
the ratio $\bar{d}(x)/\bar{u}(x)$ in the limit $x \to 0$ to be controlled 
entirely by $c_0$.  The value $c_0 = 0$ is used in the fits presented here,
so that $\bar{d}(x)/\bar{u}(x) \to 1$ at $x \to 0$.  

In future work, it 
might be preferable to include an additional factor of $(1-x)^{a_2}$ in 
$\bar{d}(x)/\bar{u}(x)$, to allow $\bar{d}(x)$ and 
$\bar{u}(x)$ to have different asymptotic behaviors at $x \to 1$.  However,
the dominance of the quark distributions by valence contributions at large 
$x$ means that this would be unlikely to affect the phenomenology.  Similarly,
one might prefer to include an additional factor of $x^{a_1}$ here, to allow 
$\bar{d}(x)$ and $\bar{u}(x)$ to have different limiting power laws at 
$x \to 0$.  However, that would involve lifting the Regge assumption that 
$\bar{d}$, $\bar{u}$, and $\bar{s}$ all have the same small-$x$ power law 
behavior.  It would also have little effect on the phenomenology, because 
we find that assuming an arbitrary constant limiting value for 
$\bar{d}(x)/\bar{u}(x)$ at $x \to 0$ results in a very large uncertainty 
in the value of that constant.

There are additional free parameters associated with the strange and sea 
quark momentum fractions and the $x^{a_1}$ and $(1-x)^{a_2}$ 
factors---with some of the $a_1$ parameters tied together 
by Regge theory.  In all, the Chebyshev fits of Sec.\ \ref{sec:CT10} 
have 84 free parameters:  $c_1,\dots,c_{12}$ in Eq.\ (\ref{eq:tseries}) 
for $u_v$, $d_v$, $g$, $\bar{d}+\bar{u}$, and $\bar{s}$; 
$c_1,\dots,c_{13}$ in Eq.\ (\ref{eq:dbaroubar}) for  $\bar{d}/\bar{u}$;
$a_2$ in Eq.\ (\ref{eq:fofx}) for $u_v$, $d_v$, $g$, $\bar{d}+\bar{u}$, $\bar{s}$;
$a_1$ in Eq.\ (\ref{eq:fofx}) for $u_v$, $g$, $\bar{d}+\bar{u}$, $\bar{s}$,  
with $u_v$ and $d_v$ values equal;
$a_0$ in Eq.\ (\ref{eq:fofx}) for $\bar{d}+\bar{u}$, $\bar{s}$,  
with $a_0$ for $u_v$ and $d_v$ determined by the number sum rules
(\ref{eq:numsum}) and for $g$ by the momentum sum rule 
(\ref{eq:momsum}).

To make a fair comparison with CTEQ6.6-style fits, the Chebyshev fits 
discussed in Sec.\ \ref{sec:fits} retained the same parametrization 
for $\bar{s}(x)$ as in CTEQ6.6 and require $\bar{d}(x)/\bar{u}(x) \to 1$ 
at $x \to 0$, leaving 71 free parameters for these fits.

The Chebyshev parametrizations used here have 3--4
times as many adjustable parameters as have been used in 
traditional PDF analyses.  This provides sufficient flexibility 
to avoid the systematic error of ``parametrization dependence,'' 
but it also means that the parametrized forms can easily take on 
more fine structure in $x$ than is plausible in the 
nonperturbative physics that is being described.  
To avoid undesirable fine structure, we adopt a strategy of 
adding a penalty to the $\chi^2$ measure 
of fit quality, based on the amount of complexity in the fitting functions.

To construct a suitable penalty, let us observe that the classic form 
\begin{equation}
f(x) = a_0 \, x^{a_1} \, (1-x)^{a_2} \; ,
\end{equation}
which surely embodies the appropriate smoothness, 
has the property that 
\begin{equation}
x \, (1-x) \, d(\ln f)/dx \, = \, a_1 - (a_1+a_2)x
\end{equation}
is linear in $x$.  Hence it is natural to define 
\begin{equation}
\Phi_a(x) \, = \, x \, (1-x) \, d(\ln f_a)/dx 
\label{eq:Phi}
\end{equation}
for each flavor $a = u$, $d$, $\bar{u}$, $\bar{d}$, $s$, and $g$.
The extent to which $\Phi_a(x)$ departs from a linear 
function is a good local measure of nonsmoothness, 
so we define
\begin{equation}
S_a \, = \, \int_{x_1}^{x_2} 
\left(\frac{d^2\Phi_a}{dx^2}\right)^2 dx \; .
\label{eq:measure}
\end{equation}

For any given set of fitting parameters, the nonsmoothness 
measure $S_a$ is computed for each of the 6 flavors at the 
input scale $\mu_0$.  These measures are multiplied by suitably 
chosen weight factors $C_a$, and the result 
\begin{equation}
\sum_a \, C_a \, S_a
\end{equation}
is added to $\chi^2$ to define the overall measure of fit 
quality that is minimized to determine best-fit parameters.  
In this initial study, the weight factors were chosen by hand 
to make the penalty term for each flavor contribute on the 
order of 1--2 to the total, so that the goodness-of-fit measure 
remained dominantly based on the traditional $\chi^2 \sim 3000$.
In detail, the values $x_1 = 0.02$ and $x_2 = 0.95$ were used, and the 
integral in Eq.~(\ref{eq:measure}) was calculated numerically 
by dividing the integration region into 200 equal bins, with 
the derivative in Eq.~(\ref{eq:Phi}) also calculated numerically. 

To enforce smoothness at small $x$, we note that the desired limiting 
behavior is given by 
\begin{equation}
f(x) = a_0 \, x^{a_1} \; ,
\end{equation}
which has the property that 
\begin{equation}
x \, d(\ln f)/dx \, = \, a_1 
\end{equation}
is constant in $x$.  Hence it is natural to define 
\begin{eqnarray}
\Psi_a(x) \, &=& \, x \, d(\ln f_a)/dx  \nonumber \\
U_a \, &=& \, \int_{\ln x_3}^{\ln x_4} 
\left(\frac{d\Psi_a}{dx}\right)^2 \, d(\ln x) \; .
\label{eq:Psi}
\end{eqnarray}
Similarly to the above, 
\begin{equation}
\sum_a \, D_a \, U_a
\end{equation}
is added to the goodness-of-fit measure, with constants 
$D_a$ chosen to make the contribution from each flavor 
$\approx 1 - 2$.  The limits used were $x_3 = 10^{-5}$ 
and $x_4 = 0.04\,$. 
(The fits discussed in Sec.\ \ref{sec:fits} and shown in Fig.\ 2 were made 
using an earlier method to enforce smoothness of the Chebyshev fits. That 
method is superceded by the method described here.)

\end{document}